\documentclass[%
    reprint,
    amsmath,
    amssymb,
    aip,
    superscriptaddress,
]{revtex4-1}

\usepackage{graphicx}		         
\usepackage{hyperref}
\usepackage[capitalise]{cleveref}
\usepackage{physics} 
\usepackage{subcaption}
\usepackage{booktabs} 
\usepackage{multirow}
\usepackage[dvipsnames]{xcolor}
\usepackage{tikz}
\usepackage[utf8]{inputenc}
\usepackage{tabularx}
\usepackage{siunitx}

\DeclareUnicodeCharacter{212B}{\AA}

\newcommand{\Markup}[1]{{\color{black}{#1}}}
\newcommand{\Markupp}[1]{{\color{black}{#1}}}

\begin{document}

\title{\Markup{Near-Exact Nuclear Gradients of Complete Active Space Self-Consistent Field Wave Functions}}

\author{\href{https://orcid.org/0000-0002-5130-8633}{James E. T. Smith}}
\affiliation{Center for Computational Quantum Physics, Flatiron Institute, New York, New York 10010 USA}
 \affiliation{Department of Chemistry, University of Colorado Boulder, Boulder, Colorado 80309, USA}
 \email{jsmith@flatironinstitute.org}

 \author{\href{https://orcid.org/0000-0002-9667-1081}{Joonho Lee}}
 \affiliation{Department of Chemistry, Columbia University, New York, New York 10027, USA}
 \email{linusjoonho@gmail.com}

 \author{\href{https://orcid.org/0000-0002-6598-8887}{Sandeep Sharma}}
 \affiliation{Department of Chemistry, University of Colorado Boulder, Boulder, Colorado 80309, USA}
 \email{sandeep.sharma@colorado.edu}


\newpage
\begin{abstract}
In this paper, we study the nuclear gradients of heat bath configuration interaction self-consistent field (HCISCF) wave functions and use them to optimize molecular geometries for various molecules.
We show that the HCISCF nuclear gradients are fairly insensitive to the size of the "selected" variational space, which allows us to reduce the computational cost without introducing significant error.
The ability of HCISCF to treat larger active spaces combined with the flexibility for users to control the computational cost makes the method very attractive for studying strongly correlated systems which require a larger active space than possible with complete active space self-consistent field (CASSCF).
Finally, we study the realistic catalyst, Fe(PDI), and highlight some of the challenges this system poses for density functional theory (DFT).
We demonstrate how HCISCF can clarify the energetic stability of geometries obtained from DFT when the results are strongly dependent on the functional.
We also use the HCISCF gradients to optimize geometries for this species and study the adiabatic singlet-triplet gap.
During geometry optimization, we find that multiple near-degenerate local minima exist on the triplet potential energy surface.
\end{abstract}

\maketitle

\section{\label{sec:Introduction}Introduction}

Systems that contain transition metals, covalent bond breaking, and electronically excited states are often challenging for theoretical methods because their electronic structure can be dominated by more than one electronic configuration.
To study such systems we require multireference approaches like the complete active space self-consistent field (CASSCF) method.\cite{Roos1980,Olsen1988,Malmqvist1990}
In CASSCF, we perform the full configuration interaction (FCI) procedure on a subset of the molecular orbitals termed the active space.
The FCI expansion enumerates all possible configurations for a given set of orbitals and scales combinatorially with the size of the active space.
This steep cost limits the size of active spaces that can be treated to roughly 22 electrons in 22 orbitals, which we abbreviate as (22e,22o).\cite{Vogiatzis2017}
A substantial amount of research in quantum chemistry is devoted to reducing the cost of configuration interaction (CI) methods through a variety of approximations:
restricted active space,\cite{Malmqvist1990,Celani2000}
generalized active space,\cite{Olsen1988,Fleig2001,Ma2011a}
density matrix renormalization,\cite{White1992,White1993,Fano1998,White1999,sch05,legeza2015,sch11,Daul2001,Chan2002, Moritz2006,Zgid2008,Luo2010,Marti2010,Chan2011,Kurashige2011,sharmaspin, sha-nat,kura-nat,Wouters2014,keller,yuki-review,Yanai2015}
selected configuration interaction (SCI),\cite{Ivanic2001,Huron1973,Buenker1974,Evangelista1983,Harrison1991,Steiner1994,Wenzel1996,Neese2003,Abrams2005,Bytautas2009,Evangelista2014,Knowles2015,Schriber2016,Liu2016,Caffarel2016,Tubman2016a,yann2017,Tubman2018,Tubman2020,Levine2020}
FCI quantum monte carlo,\cite{Booth2009,Cleland2010,Petruzielo2012,thomas2015}.
The heat bath configuration interaction (HCI) is a particularly efficient implementation of the SCI method.\cite{Holmes2016,Sharma2017,Smith2017}
These methods can correlate more than 40 electrons in 40 orbitals in routine calculations, making the study of larger and more complex systems accessible.

With the ability to treat larger active spaces, extending these algorithms to calculate molecular properties other than single-point energy is vital for connecting theoretical and experimental research.
One such property of interest is the nuclear gradients of the electronic energy which are crucial for chemical applications because they are used to obtain minimum energy geometries, transition states, reaction pathways, and are used in ab initio molecular dynamics.\cite{Curchod2018}
Many of the approximate CI/CASSCF schemes have already been extended to calculate the nuclear gradients\cite{Guareschi2013,Liu2013,Hu2015,Maradzike2017,Ma2017b,Schlimgen2018,Dash2018,Dash2019,Freitag2019,Park2021a,Park2021} and in this work we extend the family of HCI methods.

The recently-developed HCI algorithm has proved to be an efficient approximation of FCI where the size of the CI expansion is controlled with a single user parameter $\epsilon_1$.
This parameter allows one to continuously grow the wave function from a single determinant, e.g. Hatree-Fock (HF), to all configurations in the FCI expansion.
The energy of this wave function can be corrected with a perturbative correction which is often necessary to reduce the error, relative to FCI, to an acceptable level. 
In recent work, two of the authors extended the HCI algorithm to be used in CASSCF-like calculations and implemented this method in the open source quantum chemistry package \texttt{PySCF}.\cite{Sun2015,Sun2017,Sun2018,Sun2020}
This work showed that the multireference orbitals can be obtained from HCISCF with relatively loose thresholds and as long as a final step in which the optimized orbitals are used to perform a single tight threshold HCI calculation, the final results are in good agreement with full CASSCF calculation.\cite{Smith2017}
In practice, this insensitivity of the final results to the optimized orbitals is advantageous because it means that accurate multireference orbitals can be obtained from relatively cheap HCI calculations.
In general, other properties may be also be insensitive to the accuracy of the HCI wave function, which motivates this work and the study of nuclear gradients.

\Markup{
Previous work on approximate CI/CASSCF schemes has shown that for most systems, properties such as gradients and relaxed geometries are often quite accurate even when the energy is not.\cite{Guareschi2013,Liu2013,Hu2015,Maradzike2017,Ma2017b,Schlimgen2018,Dash2018,Dash2019,Freitag2019,Park2021a,Park2021}
In particular, Park has studied the gradients of a different selected CI scheme, adaptive sampling CI (ASCI), and demonstrated that CASSCF gradients and geometries were well approximated by this method.\cite{Park2021a,Park2021}
Park showed that geometries converged faster than energies with respect to the number of determinants in the wave function and that the convergence of both energy and relaxed geometries could be improved by a second order perturbation theory correction (PT2) and by extrapolating these corrections.\cite{Park2021} As a result, important properties of these molecules like singlet-triplet (S-T) gaps were also improvable with these corrections, but we note that for most systems, the difference after the correction was typically around 1 kcal/mol.
However, Park's work focused largely on aromatic systems, which are not particularly strongly correlated, while in this work we focus on the transition metal complex Fe(PDI), which is highly multireference.\cite{Ortuno2017} 
}

The rest of this paper is organized as follows.
First, we briefly review HCI theory and its self-consistent variant along with analytical gradients of CI wave functions.
Next, we discuss the gradients of CASSCF-like wave functions where HCI is used as the CI solver and verify that they converge smoothly with the HCI parameter $\epsilon_1$.
Then, we highlight several strategies to improve the quality of the gradients when increasing the size of the CI expansion is not feasible, e.g. when memory is limited.
Finally, we demonstrate the use of these gradients to find equilibrium geometries and adiabatic singlet-triplet gaps for the model catalytic system (PDI)Fe-N$_2$, where PDI is 2,6-bis[1-(2,6-dimethylphenyl-imino)ethyl]pyridine.
For simplicity we refer to this complex as Fe(PDI).
For this challenging system, we also examine the important practical consideration of selecting the active space in a robust numerical manner.

\begin{figure*}
    \centering
    
    \begin{tikzpicture}
    \node at (0,0) {\includegraphics[width=\textwidth]{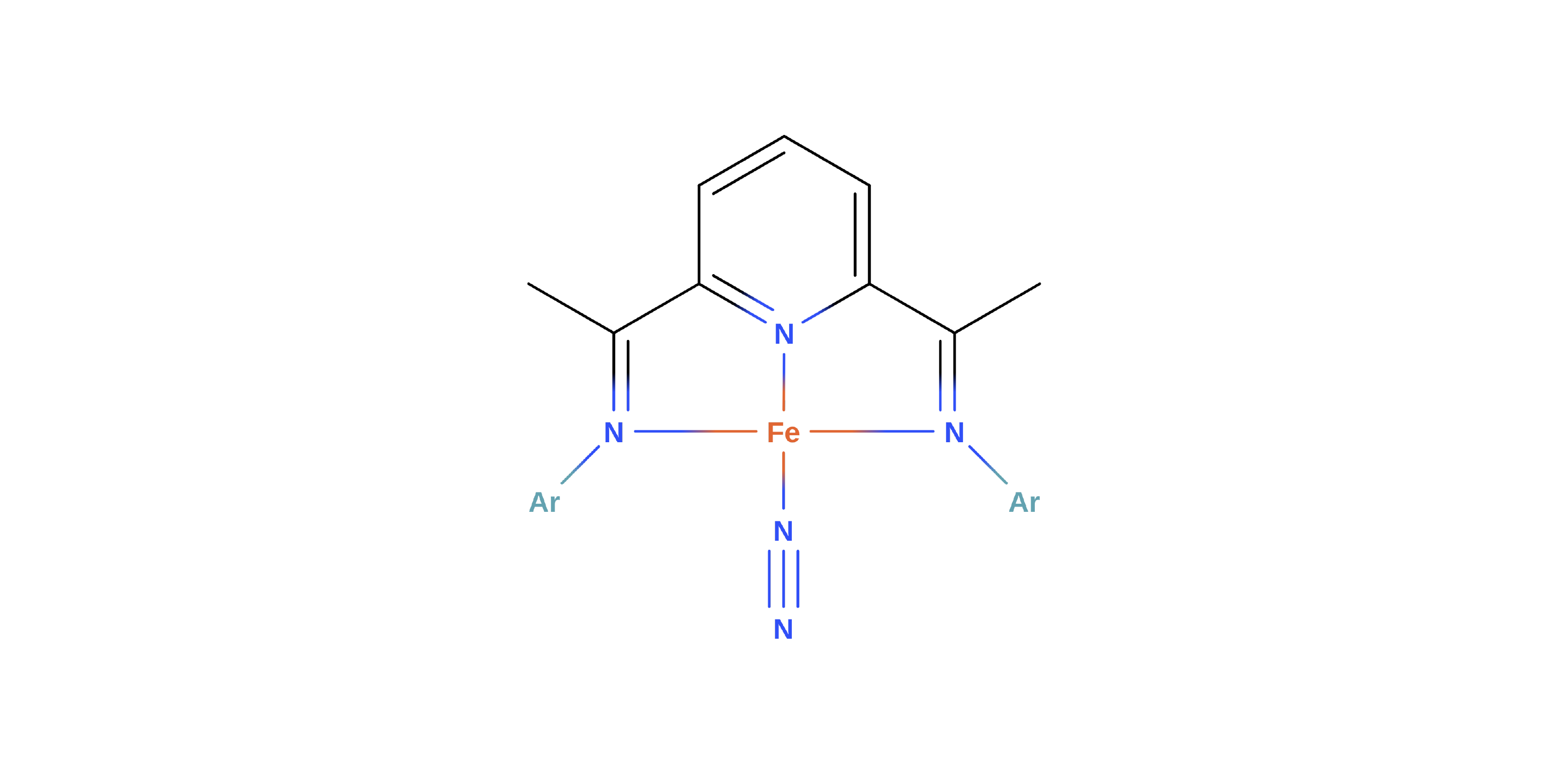}};

    \node at (-2.3,   -0.125) {N1};
    \node at (0,   1.25) {N2};
    \node at (2.3,    -0.125) {N3};
    \node at (0.5, -1.5) {N4};
    \node at (0.5,   -3) {N5};
    
    \node at (0.5,0.125) {Fe1};
    
    \node at (-2, 1) {C2};
    \node at (-1.25, 1.25) {C3};
    \node at (1.25,  1.25) {C7};
    \node at (2,  1) {C8};

    \end{tikzpicture}
    \caption{The model Fe(PDI) complex. Ar = 2,6-dimethylphenyl. The numbering convention is the same as Stieber et al.\cite{Stieber2012} for consistency.}
    \label{fig:fe_pdi}
\end{figure*}
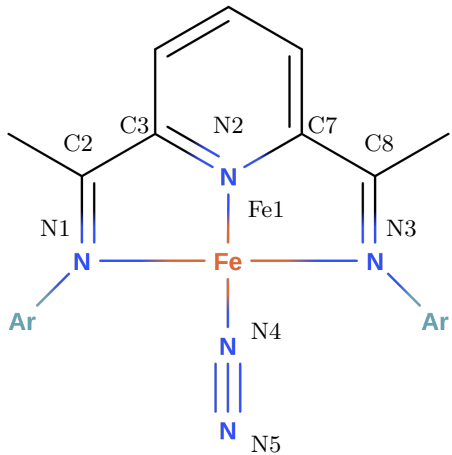

\section{\label{sec:theory}Theory}
\subsection{Heat-bath Configuration Interaction Algorithm}
HCI is a two-step procedure consisting of a variational calculation and a perturbative correction.
In the first stage of the HCI algorithm, a set of \textit{important} determinants is iteratively built and used to expand the variational wave function, then a second-order correction to the energy is computed using multireference Epstein-Nesbet perturbation theory (PT).\cite{Epstein1926,Nesbet1955}
Like other selected CI (SCI) + PT methods, it derives from FCI and linearly parametrizes the wave function $\ket{\Psi} = \sum_{i} c_i \ket{D_i}$.
Unlike FCI, where the expansion is a set all possible determinants, SCI methods express the wave function as a set of \textit{important} determinants.
This allows SCI+PT methods to correlate much larger systems than possible with FCI.

\subsubsection{Variational Stage}
In the first step, the HCI algorithm identifies a set of \textit{important} determinants that approximate the full many-body wave function and we refer to this set as the variational space ($\mathcal{V}$).
At the beginning of the variational step, the user specifies a small number of determinants to add to $\mathcal{V}$ and usually this includes only the Hartree-Fock determinant.
Then the algorithm adds all connected determinants $\ket{D_a}$ which satisfy the following importance criteria
\begin{equation}
    \max_{ \ket{D_i} \in \mathcal{V}} \left|H_{ai} c_i\right| > \epsilon_1
    \label{eqn:hci_selection}
\end{equation}
where $\ket{D_i}$ is a determinant already in the variational space, $H_{ai}=\bra{D_a}\hat{H}\ket{D_i}$, $c_i$ is the amplitude of the $\ket{D_i}$, and $\epsilon_1$ is a user-specified parameter that controls the size of the variational space, i.e. the accuracy of the variational wave function.
A smaller $\epsilon_1$ means that the $\mathcal{V}$ is closer to the FCI determinant space and in the limit that $\epsilon_1$ goes to 0, we recover our FCI wave function and energy.
We repeat this selection of important determinants and grow the variational space until the HCI energy converges.

The CIPSI selection criteria, $ \frac{\sum_i H_{ki}c_i}{E_0 - H_{kk}} > \epsilon_{\text{CIPSI}}$, is more complicated and uses first order perturbation theory to determine whether a determinant should be added to the variational space.
Despite its similarity to the CIPSI selection criteria, the HCI importance function offers several key advantages over the \textit{original} CIPSI algorithm.
Since the matrix elements between determinants connected by double-excitations are just the two-electron integral elements (and a parity factor of +1 or -1), evaluating the HCI selection criteria requires no calculations.
While avoiding calculations is advantageous, the real gain in performance comes from the fact that the two electron integrals are stored and sorted \textit{once} before the variational stage.
Selecting \textit{important} determinants only requires iterating through the sorted two-electrons integrals and the algorithm stops once those values fall below a certain threshold.
As a result, this means that the algorithm never even "touches" the unimportant connected determinants, which often outnumber the \textit{important} ones by many orders of magnitude.
We note that HCI was inspired by challenges in the original formulation of CIPSI and more modern implementations have shown substantial improvements compared to their predecessors.\cite{Garniron2019}

\subsubsection{Perturbative Stage}
We then correct the variational energy using a second order correction from Epstein-Nesbet perturbation theory, evaluated as
\begin{equation}
    \label{eqn:enpt}
    E_2 (\epsilon_2)
    =
    \sum_{\ket{D_a}} 
    \Bigg(
    \sum_{\ket{D_i} \in \mathcal{V}}^{(\epsilon_2)} H_{ai}c_i
    \Bigg)^2
\end{equation}
where the inner summation is "screened" in a similar manner to the variational stage.
Analogously, the perturbative expansion grows as $\epsilon_2$ becomes smaller and in the limit $\epsilon_2\rightarrow0$ the perturbative correction becomes exact.

While screening the inner sum in \cref{eqn:enpt} mitigates the large memory requirement for calculating a perturbative correction it does not eliminate it entirely.
Calculating an accurate correction requires a small $\epsilon_2$ value and for large systems, evaluating \cref{eqn:enpt} deterministically becomes prohibitively expensive.
In these cases, we can reduce the memory requirement further by calculating $E_2(\epsilon_2)$ semi-stochastically.
By that we mean that we calculate as much of the correction deterministically as computational resources allow and then calculate the remainder stochastically.
This procedure is advantageous compared to a purely stochastic one because the portion of $E_2(\epsilon_2)$ calculated stochastically is smaller and it takes fewer iterations to converge the stochastic error.
We refer to calculations that use this procedure as semi-stochastic HCI (SHCI) to distinguish them from HCI calculations that only use the variational step.
There is another even more efficient scheme, which was recently introduced by one of the authors of this paper, but we do not use that scheme in this work.\cite{Li2018a}

\subsection{\label{subsec:hci_grad:grad_theory}Gradients of HCISCF Wave Functions}
We briefly review analytical energy gradients of variational wave functions.
For a detailed discussion on the subject, we refer the reader to the review by Yamaguchi and Schaefer.\cite{Yamaguchi2011}
We start with the electronic Hamiltonian in second quantization:

\begin{equation}
    \hat{H} =
    \sum_{ij\sigma} h_{ij} a^\dagger_{i\sigma} a_{j\sigma}
    +
    \sum_{ijkl\sigma\sigma'} g_{ijkl} a^\dagger_{i\sigma} a^\dagger_{j\sigma'} a_{k\sigma'} a_{l\sigma}
\end{equation}
We define the one- and two-body integrals in the usual way: 
\begin{equation}
    \begin{split}
        h_{ij} &= \int \phi_i^*(\mathbf{x}) \Big( 
        - \frac{1}{2} \nabla^2 - \sum_I \frac{Z_I}{r_I}
        \Big)
        \phi_j(\mathbf{x}) d\mathbf{x}\\
        g_{ijkl} &= \int \int 
        \frac{\phi_i^*(\mathbf{x_1}) \phi_j^*(\mathbf{x_2}) \phi_k(\mathbf{x_1}) \phi_l(\mathbf{x_2})}{r_{12}}
        d\mathbf{x_1} d\mathbf{x_2}
    \end{split}
\end{equation}
where $\phi$ are orbitals, $Z_I$ are the nuclear charges, $r_I$ the electron-nuclear separations, and $r_{12}$ the electron electron separation.
We use $i$, $j$, $k$, and $l$ as orthogonal orbital indices and $\sigma$, $\sigma'$ as spin indices.
In most cases, these are molecular orbitals (MOs), which are linear combinations of atomic orbitals (AOs) $\ket{i} = \sum_\mu^{AO} C_\mu^i \ket{\mu}$.
For the remainder of this work, we use the following convention for orbital indices: Greek letters, e.g. $\mu$ and $\nu$ refer to AOs, while $i$, $j$ refer to MOs.

The active space Hamiltonian is a function of the basis set, the MO coefficients $\mathbf{C}$, and the nuclear coordinates $\{a_i\}$ while
the electronic wave function $\ket{\Psi}$ is a function of the parameters $\{c_i\}$ making the electronic energy a function of all four.
Expanding the nuclear gradients in terms of these variables we get:

\begin{equation}
    \label{eqn:energy_grad_general}
    \frac{dE_{elec}}{d a} = 
    \frac{\partial E_{elec}}{\partial a} + 
    \sum_{i\mu} \frac{\partial E_{elec}}{\partial C_\mu^i} \frac{d C_\mu^i}{d a} + 
    \sum_i \frac{\partial E_{elec}}{\partial c_i} \frac{d c_i}{d a}
\end{equation}

The first term is the contribution from the basis set, i.e. the atomic orbitals, the second term is from the MO coefficients, and the final term is from the wave function parameters.
\Markup{
Since the CI coefficients are always determined variationally so $\frac{\partial E_{elec}}{\partial c_i}=0$ the term always vanishes.
Similarly, for wave functions where the molecular orbital parameters are variational, $\frac{\partial E_{elec}}{\partial C_\mu^i}=0$ and the second term on the right hand side of \cref{eqn:energy_grad_general} also vanishes.
}
To derive an explicit form for the gradients, we start by expressing the electronic energy in second-quantized form:

\begin{equation}
    \label{eqn:energy_second_quant}
    E_{elec} =
    \sum_{ij} h_{ij} \gamma_{ij} +
    \sum_{ijkl} g_{ijkl} \Gamma_{ijkl}
\end{equation}
where $\gamma_{ij} = \sum_\sigma \bra{\Psi} a^\dagger_{i\sigma} a_{j\sigma} \ket{\Psi}$ is the one-body reduced density matrix (RDM)
and $\Gamma_{ijkl} = \frac{1}{2} \sum_{\sigma \sigma'} \bra{\Psi} a^\dagger_{i\sigma} a^\dagger_{j\sigma'} a_{k\sigma'} a_{l\sigma} \ket{\Psi}$
is the two-body RDM.
Taking derivative of \cref{eqn:energy_second_quant} with respect to a general nuclear coordinate $a$ generates terms including $\frac{\partial \gamma_{ij}}{\partial a}$ and $\frac{\partial \Gamma{ijkl}}{\partial a}$.
Since the RDMs are just functions of the CI coefficients, we know we can ignore their derivatives because the final term in \cref{eqn:energy_grad_general} vanishes.
As a result, we arrive at a concise expression for then the nuclear gradients of the energy:

\begin{equation}
    \label{eqn:ci_grad}
    \frac{d E_{elec}}{d a} =
    \sum_{ij} \frac{d h_{ij}}{d a} \gamma_{ij} +
    \sum_{ijkl} \frac{d g_{ijkl}}{d a} \Gamma_{ijkl}
\end{equation}

Expanding this equation further, we can separate the terms relating to the MO response:

\begin{equation}
    \label{eqn:ci_grad_expanded}
\begin{split}
    \frac{d E_{elec}}{d a} &=
    \sum_{ij} h_{ij}^a \gamma_{ij} + \sum_{ijkl} g_{ijkl}^a \Gamma_{ijkl} \\
    &+
    \sum_{ij} U_{ij}^a (X_{ij}-X_{ji})
    - 
    \sum_{ij} S_{ij}^a X_{ij}
\end{split}
\end{equation}
Where $h_{ij}^a$, $g_{ijkl}^a$, and $S^a_{ij}$ are the "skeleton" derivatives of the one-electron, two-electron and overlap integrals defined as 
\begin{equation}
    \begin{split}
        h_{ij}^a &= \sum_{\mu\nu}^{AO} C_\mu^i C_\nu^j \frac{d h_{\mu\nu}}{d a} \\
        g_{ijkl}^a &= \sum_{\mu\nu\rho\sigma}^{AO} C_\mu^i C_\nu^j C_\rho^k C_\sigma^l\frac{d g_{\mu\nu\rho\sigma}}{d a} \\
        S_{ij}^a &= \sum_{\mu\nu}^{AO} C_\mu^i C_\nu^j \frac{d S_{\mu\nu}}{d a} \\
    \end{split}
\end{equation}
$U_{ij}^a$ is called the orbital response and is defined by 
\begin{equation}
    \frac{d C_\mu^i}{d a} = \sum_m^{MO} U_{mi}^aC_\mu^m
\end{equation} 
It can be determined by solving the z-vector equations.\cite{Handy1984}
Finally, $X_{ij}$ is the Lagrangian from the Generalized Brillouin Theorem, defined as 
\begin{equation}
    X_{ij} = \sum_{m} h_{im} \gamma_{jm} + 2\sum_{mkl} g_{imkl} \Gamma_{jmkl}
\end{equation}

For CASSCF, $X_{ij}$ is symmetric, i.e. $X_{ij} = X_{ji}$, and we can simplify \cref{eqn:ci_grad_expanded} to 
\begin{equation}
    \label{eqn:casscf_grad}
    \frac{d E_{elec}}{d a} =
    \sum_{ij} h_{ij}^a \gamma_{ij} + \sum_{ijkl} g_{ijkl}^a \Gamma_{ijkl}
    -
    \sum_{ij} S_{ij}^a X_{ij}
\end{equation}

and avoid solving the z-vector equations.
We direct the reader to the derivation of Equation (133) in the excellent review by Yamaguchi and Schaefer.\cite{Yamaguchi2011}
All three terms in \cref{eqn:casscf_grad} arise from response of the basis set to changes in nuclear coordinates and no terms from the response of MO or CI coefficients are present. 

This derivation hinges on the fact that $X_{ij}$ is symmetric and for approximate CASSCF schemes like HCISCF and DMRGSCF, $X_{ij}$ is generally not symmetric and since we ignore the third term in \cref{eqn:casscf_grad}, it is not exact.
In practice, one can solve include rotation among the active space orbitals, which we refer to as active-active (AA) rotations during the orbital optimization to eliminate the error of using \cref{eqn:casscf_grad} with approximate schemes like DMRGSCF and HCISCF.
Despite the formal lack of exactness, we highlight that if these approximate CI methods are sufficiently converged, the expression for the gradients is a justified approximation.
If these methods are not converged to the FCI limit, \cref{eqn:casscf_grad} must include a response term due to the AA rotations to be formally exact.

\Markup{
Finally, we note that the screening in \cref{eqn:hci_selection} and \cref{eqn:enpt} are dependent on the geometry, meaning that the potential energy surface can be discontinuous for a fixed value of the $\epsilon_1$.
During the finite-difference calculations used to verify the analytic gradients, we only observed discontinuities for extremely inaccurate HCI wave functions where the HCI parameter $\epsilon_1$ is greater than 10 mHa and the wave functions consisted of less than 100 determinants.
Since typical HCI calculations use much smaller values of $\epsilon_1$ and many more determinants, we do not explore these discontinuities further in this work.
}

\section{\label{comp_details}Computational Details}

For all DFT geometry optimizations, we use \texttt{Gaussian16} with the def2-TZVP basis set and density fitting to reduce the computational cost.\cite{Frisch2016} We use the default convergence thresholds in the \texttt{Gaussian16} package.
In addition to finding optimal geometries, we used harmonic frequency analysis to confirm that the reported geometries are all genuine local minima, i.e. contain no imaginary frequencies.
We perform all other quantum chemical calculations using \texttt{PySCF}\cite{Sun2015,Sun2017,Sun2018,Sun2020} and checked several of the mean-field calculations using \texttt{Q-Chem}.\cite{Epifanovsky2021}
For HCISCF calculations we use our own implementation of the HCI algorithm, called \texttt{Dice},\cite{Smith2017} in tandem with \texttt{PySCF}.
The nuclear gradients are calculated in \texttt{PySCF} using the RDMs obtained from \texttt{Dice} and \texttt{geomeTRIC}\cite{Wang2016} is used to perform the geometry updates.
Due to the challenging nature of Fe(PDI) problem we relaxed the geometry convergence thresholds for all optimizations of the Fe(PDI) species, which are each double the default thresholds used by both \texttt{Gaussian16} and \texttt{geomeTRIC}.
Our convergence thresholds are: 
$2\cdot10^{-6}$ Ha for the energy $(\tau_E)$, 
$6\cdot10^{-4}$ Ha/Bohr for the root mean squared (RMS) of the nuclear gradient $(\tau_{grms})$, 
$9\cdot10^{-4}$ Ha/Bohr for the maximum of the nuclear gradient $(\tau_{gmax})$, 
$2.4\cdot10^{-3}$ \AA \text{} for the RMS of the atomic displacement $(\tau_{drms})$, and
$3.6\cdot10^{-3}$ \AA \text{} for the maximum atomic displacement $(\tau_{dmax})$.


For the Fe(PDI) species, we use a numerical procedure to select the active space to reduce the heuristics which typically plague active space calculations.
All multireference calculations use the cc-pVDZ basis set.\cite{Dunning1989}.
We outline the procedure for selecting the active space orbitals below:

\begin{enumerate}
    \item \label{alg:uno:UKS} Run unrestricted DFT calculation 
    using the PBE0 functional\cite{Perdew1996,Adamo1999} to generate an initial guess for Step \ref{alg:uno:UHF}. 
    \item \label{alg:uno:UHF} Run an unrestricted Hartree-Fock (UHF) calculation to generate spin orbitals without correlation from an exchange-correlation functional.
    \item Use the procedure outlined in Ref. \citenum{Keller2015} to generate unrestricted natural orbitals (UNOs).\cite{Bofill1989}
    \item \label{alg:uno:UHF_UNO} Use the UHF natural orbital occupation numbers (NOONs) along with the spin density information from the UHF calculation to ensure that the targeted state shows strong correlation and qualitatively correct spin density.
    \item \label{alg:uno:HCISCF} Use the UNOs in a "loose" HCI calculation where the $\epsilon_1=10^{-4}$ Ha and the active space is (100e,100o). We select the orbitals based on their NOON so the 50 orbitals below the highest occupied natural orbital (HONO) and 50 above the lowest unoccupied natural orbital (LUNO) are part of the active space.
    \item Use the 1-RDM from the previous step to calculate new NOs and sort based on NOONs.
    \item Use these orbitals in all subsequent HCISCF calculations, selecting the orbitals solely based on their NOONs analogous to the way used in Step \ref{alg:uno:HCISCF}. We select the orbitals in pairs (i.e. one occupied and one virtual) around the HONO-LUNO gap. For all Fe(PDI) calculations, we use an active space of (40e,40o), which includes more than those recommended by the criteria UNO methods, i.e. $0.02 <  \text{NOON} < 1.98$.\cite{Bofill1989} 
\end{enumerate}

We emphasize that Steps \ref{alg:uno:UKS} and \ref{alg:uno:UHF} are sensitive to the optimization strategy used for the SCF procedure, e.g. DIIS,\cite{Pulay1980} ADIIS,\cite{Hu2010} or Newton's Method.
We surveyed a range of choices for these strategies in both steps and select the best pair of strategies for each species.
By ``best" here, we mean that the SCF solution is the lowest in energy, stable with respect to its variational parameters, and has non-negligible spin-density on the Fe atom.
All results from this survey are shown and described in \cref{app:uhf_survey}.
While we have tried to search thoroughly for the lowest energy UHF solution, we note that it is possible that even lower energy solutions exist in this complicated energy landscape.
From our experience these different SCF solutions could yield different multireference orbitals, i.e. distinct multiconfigurational self consistent field (MCSCF) wave functions, and as a result different optimized molecular structures.

For all geometry comparisons we use Kabsch's algorithm\cite{Kabsch1976} to align geometries and report the difference in geometry as $\frac{||\Delta||_2}{\sqrt{n_{a}}}$ where $\Delta$ is the difference in aligned nuclear coordinates and $n_a$ is the number of atoms.
For brevity, we refer to this quantity as RMSD for the remainder of the paper.
To align the geometries we use the open source package \texttt{rmsd}.\cite{Kromann2020}

The input scripts necessary to reproduce \textbf{all} of the results from this work are publicly available on GitHub at \href{https://github.com/jamesETsmith/hci_nuc_gradients}{https://github.com/jamesETsmith/hci\_nuc\_gradients}.

\section{\label{sec:results}Results}
\subsection{\label{subsec:gradients}HCISCF Gradients}

\begin{figure}[!htbp]
    \begin{subfigure}[t]{0.49\textwidth}
        \centering
        \includegraphics[width=\textwidth,trim={0 10cm 0 10cm},clip]{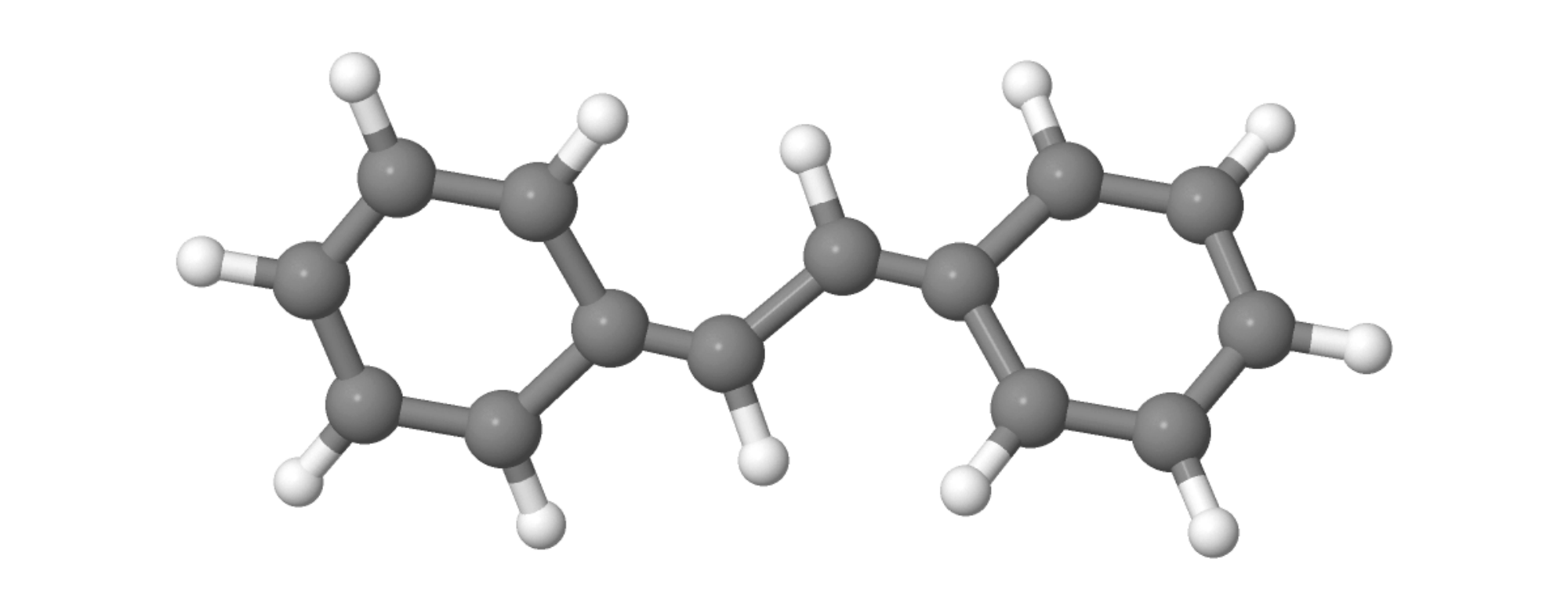}
        \caption{}
    \end{subfigure}
    \begin{subfigure}[t]{0.49\textwidth}
        \centering
        \includegraphics[width=\textwidth,trim={0 10cm 0 8cm},clip]{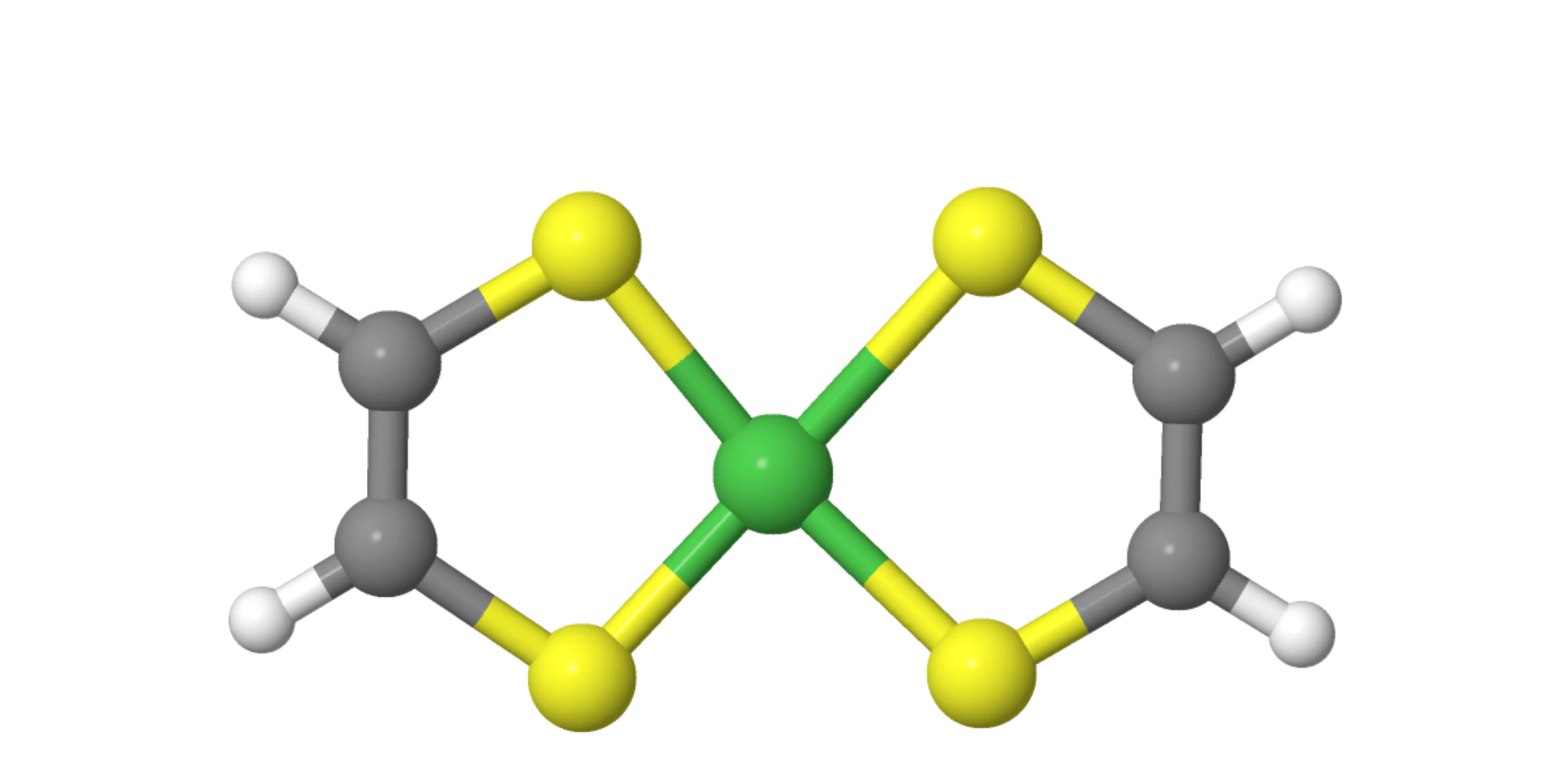}
        \caption{}
    \end{subfigure}
    \caption{The molecular systems used \cref{subsec:gradients} to study the gradients of HCI-type wave functions. (a) Stilbene and (b) bis-(ethylene-1,2-dithiolato)nickel (Ni(edt)$_2$). In these structures, gray corresponds to carbon, yellow to sulphur, green to nickel, and white to hydrogen.}
    \label{fig:eps1_conv}
\end{figure}

\begin{figure*}[!htbp]
    \begin{subfigure}[t]{0.49\textwidth}
        \centering
        \includegraphics[width=\textwidth,trim={0 0 0 1cm},clip]{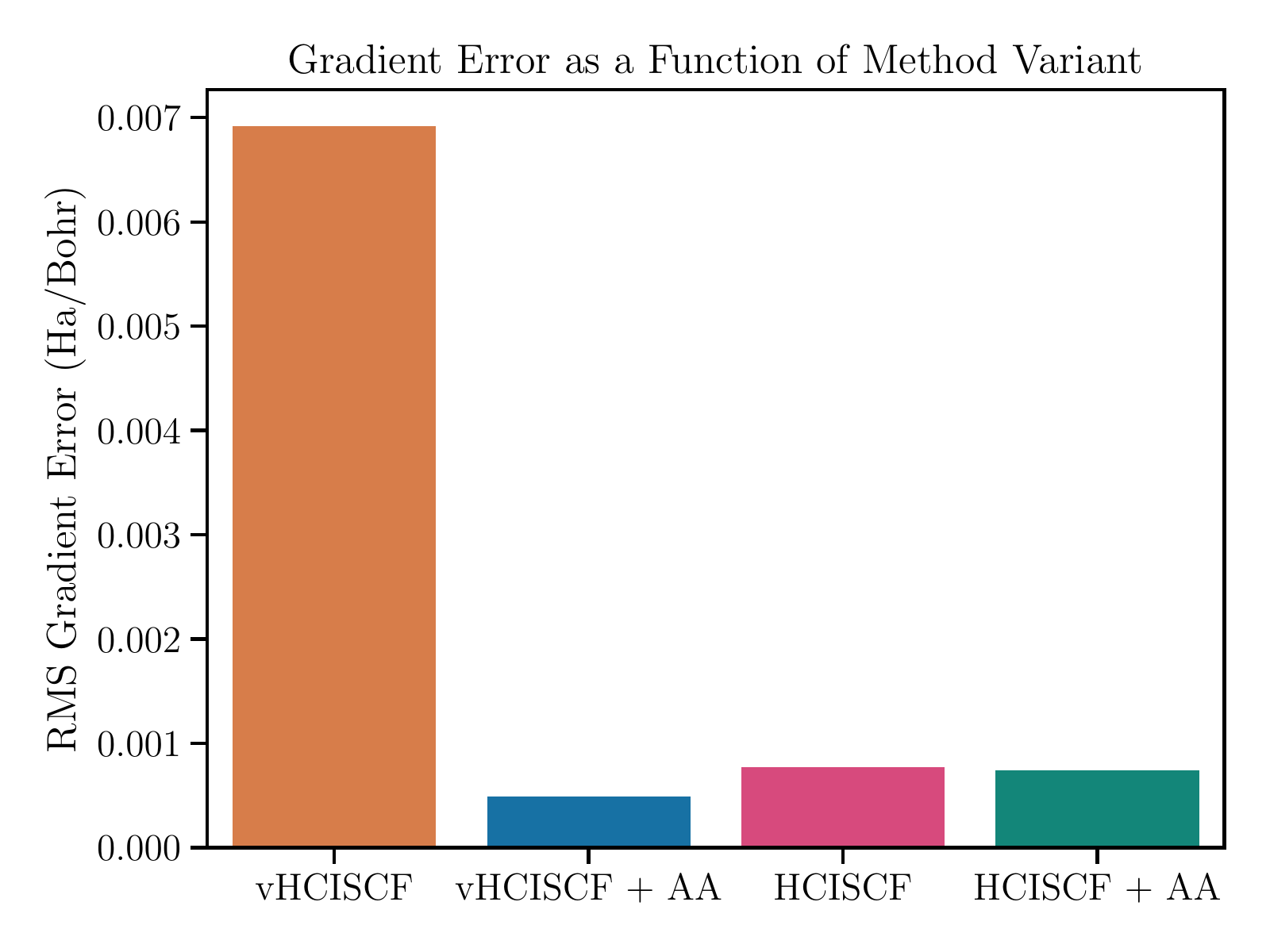}
        \caption{}
    \end{subfigure}
    \begin{subfigure}[t]{0.49\textwidth}
        \centering
        \includegraphics[width=\textwidth,trim={0 0 0 1cm},clip]{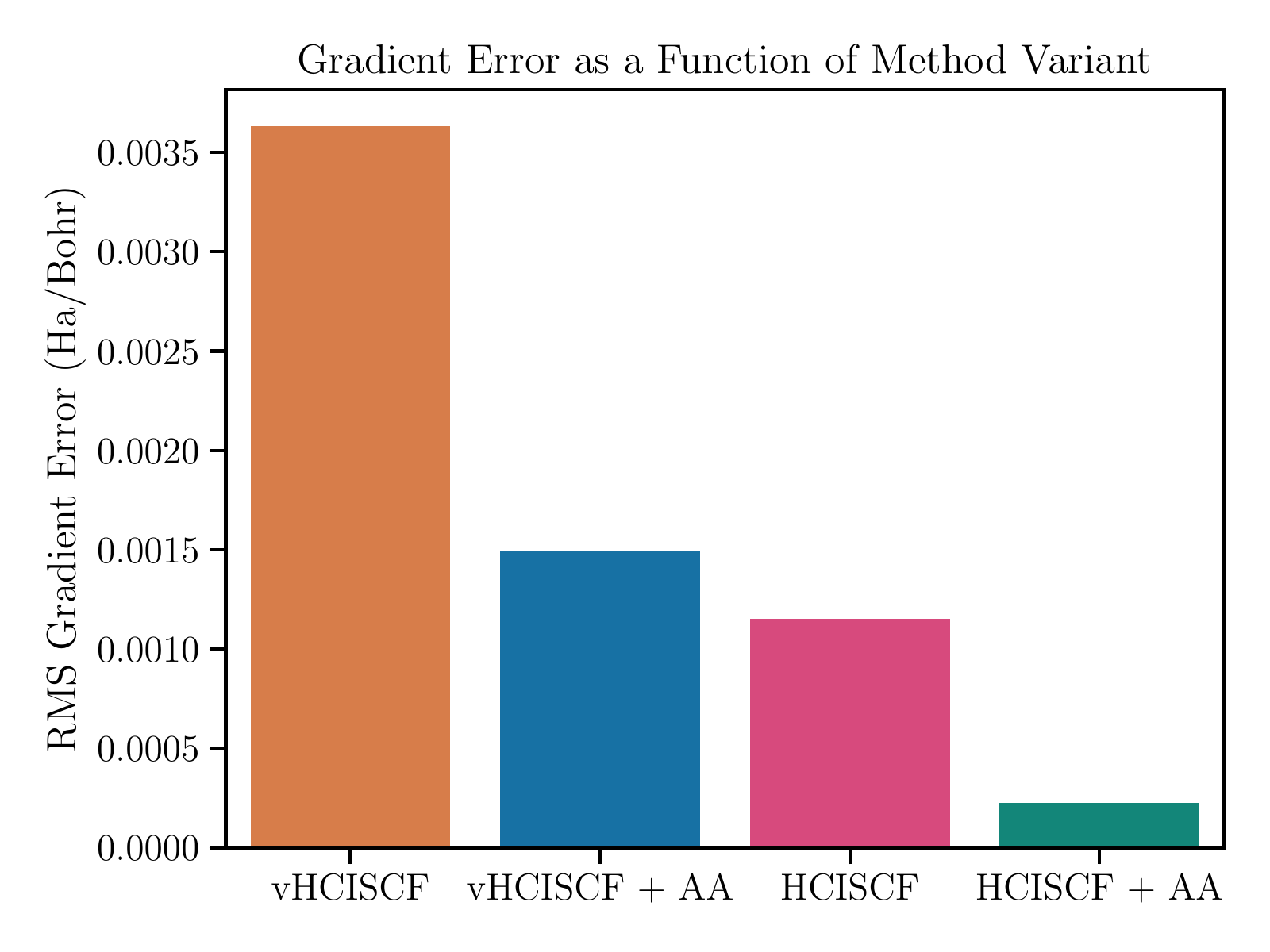}
        \caption{}
    \end{subfigure}
    \begin{subfigure}[t]{0.49\textwidth}
        \centering
        \includegraphics[width=\textwidth,trim={0 0 0 1cm},clip]{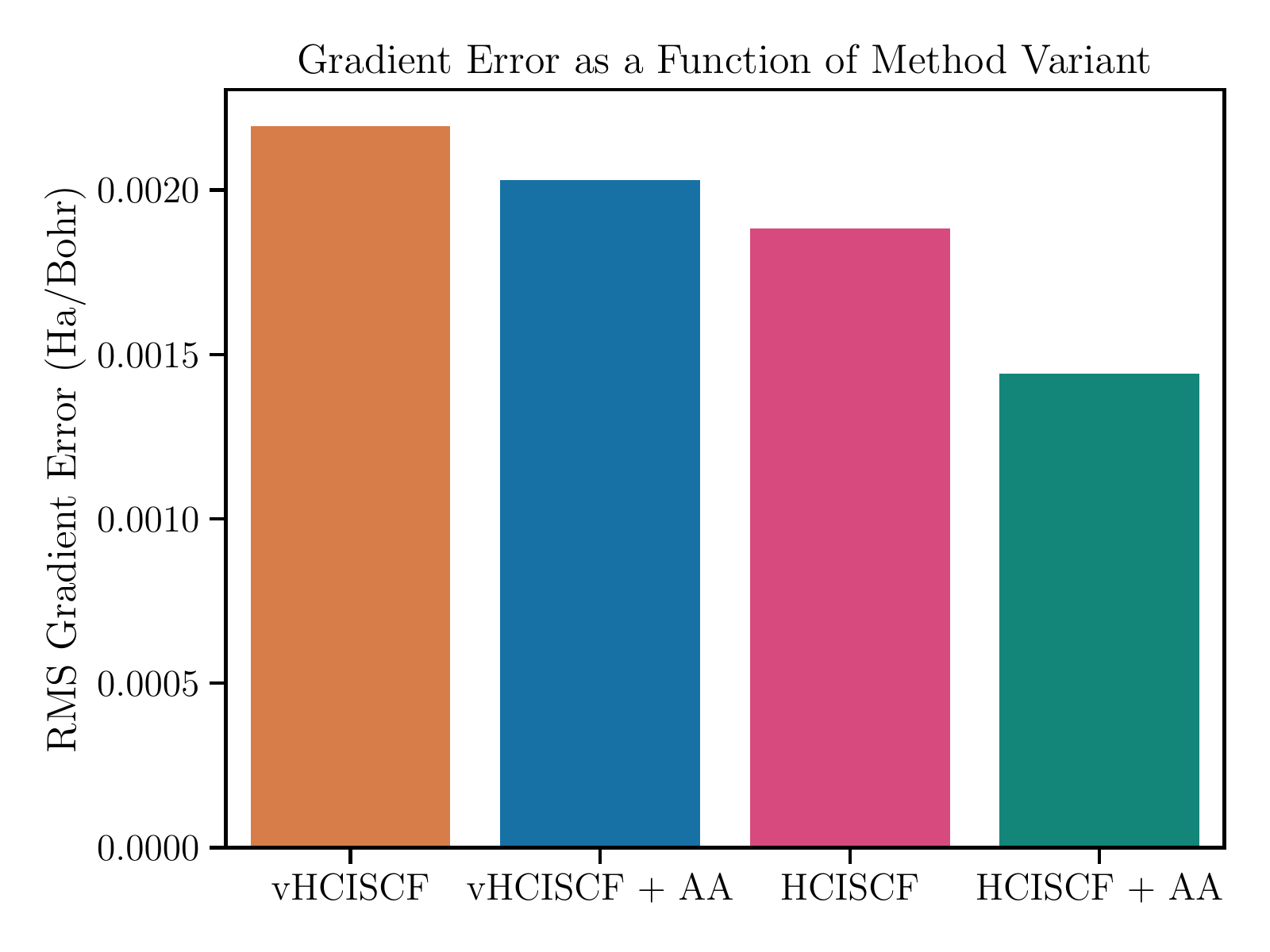}
        \caption{}
    \end{subfigure}
    \begin{subfigure}[t]{0.49\textwidth}
        \centering
        \includegraphics[width=\textwidth,trim={0 0 0 1cm},clip]{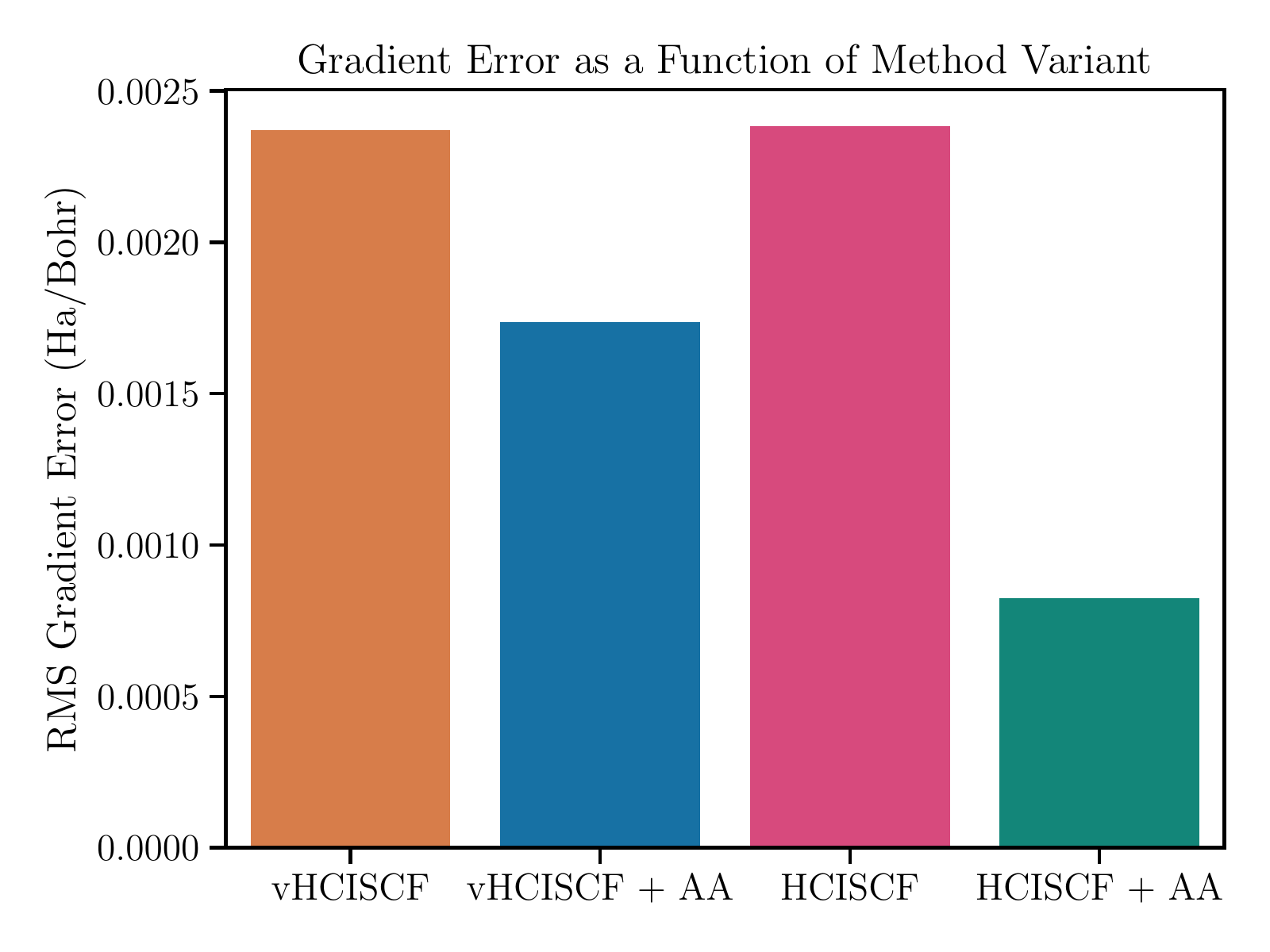}
        \caption{}
    \end{subfigure}
    \caption[Gradient errors for HCISCF wave functions.]{\Markup{Errors in gradient relative to CASSCF for various flavors of HCI-based wave functions for (a) N$_2$ CAS(10e,8o)/cc-pVDZ at a bond length of 1 \AA \text{} and (b) Sc$_2$ CAS(6e,18o)/cc-pVDZ at 2.38 \AA \text{} (c) trans-stilbene CAS(14e,14o)/cc-pVDZ and (d) Ni(edt)$_2$ CAS(18e,13o)/cc-pVDZ. For (a) and (b) used values of $\epsilon_1=5\cdot10^{-3}$ and $\epsilon_2=10^{-10}$ Ha and calculations for (c) and (d) used values of $\epsilon_1=1\cdot10^{-3}$ and $\epsilon_2=10^{-12}$ Ha. The error in vHCISCF here is high due to the deliberately large $\epsilon_1$ parameter. This was done to highlight the ability of AA rotations and the perturbative correction to effectively reduce the overall error in the gradients.}}
    \label{fig:grad_single_pt}
\end{figure*}

In this subsection, we discuss the errors that arise when using \cref{eqn:casscf_grad} for approximate CASSCF wave functions.
In \cref{fig:grad_single_pt}, we compare different variants of HCISCF gradients to CASSCF gradients, all using the cc-pVDZ basis set.\cite{Dunning1989}
As mentioned earlier, the HCISCF gradients approximate CASSCF gradients because HCI is an approximation of FCI and as a result, the energy is not invariant for AA orbital rotation.
In our discussions of gradients, we note that all gradients discussed suffer from basis set incompleteness errors, but since that is constant throughout the calculations we do not discuss it further.

Without solving the response term, there are two ways we can eliminate the error of this approximation: we can include AA rotations in the MCSCF optimization or we can converge the HCI wave function to the FCI limit.
Including the AA rotations in the MCSCF optimization removes the need for the z-vector equation completely and ensures that \cref{eqn:casscf_grad} is exact.
However it can significantly increase the number of MCSCF iterations required to reach convergence.
AA rotations make the MCSCF procedure more challenging because the corresponding parameters are strongly coupled to the CI coefficients.
Due to these challenges, optimizing AA rotations is often the most expensive way to improve the gradients.
In the second approach, we can either increase the size of the variational space or include a perturbative correction and increase the size of the perturbative space to better converge an HCI wave function.
By better converging to the FCI limit, we eliminate any effect of AA rotations on the energy, and as a result any need for the z-vector equation, again ensuring that \cref{eqn:casscf_grad} is exact. 
It is more cost-efficient to increase the number of configurations treated perturbatively and if we are memory limited this may be the only option.
When performing CASSCF calculations using only the variational step of HCI, we refer to these as vHCISCF wave functions and when we use both the variational and the pertubative, we refer to them as HCISCF.
\Markup{
In \cref{fig:grad_single_pt}, we compare the different flavors of HCI-based gradients for four systems: N$_2$, Sc$_2$, trans-stilbene, and bis-(ethylene-1,2-dithiolato)nickel (Ni(edt)$_2$) using actives spaces of (10e,8o), (6e,18o), (14e,14o), and (18e,13o) respectively.
For N$_2$ and Sc$_2$ these active space sizes corresponds to the full valence space.
These active spaces correspond to the full valence space for N$_2$ and Sc$_2$, 
the conjugated 2$p_z$ orbitals on carbon for trans-stilbene, and the conjugated 2$p_z$ orbitals of carbon and sulfur as well as the 3$d$ orbitals of the nickel atom for Ni(edt)$_2$.
Schlimgen and Mazziotti have shown that Ni(edt)$_2$ exhibits strong multireference character making it an ideal candidate to study gradients of approximate CASSCF wave functions.\cite{Schlimgen2017,Schlimgen2018}
The geometries for all systems studied are in the publicly available Github repository \href{https://github.com/jamesETsmith/hci_nuc_gradients}{https://github.com/jamesETsmith/hci\_nuc\_gradients}.
}
We start with an intentionally inaccurate vHCISCF wave function, to leave considerable room for improvement, and show that adding AA rotations or PT correction can decrease the gradient error.
\Markupp{As expected, we find that AA rotations and the PT correction improve the quality of gradients and with the exception of N$_2$ provide the largest reduction of error when used together.}

Although AA rotations and the PT correction in our MCSCF optimization can reduce the error in our nuclear gradients, they do so at a cost that is often difficult to predict.
\Markup{Both options increase the number of MCSCF iterations required for convergence and with AA rotations the calculations can require more than 50 iterations and may not converge at all.
This behavior is consistent with previous DRMG-SCF\cite{Ma2017a} and ASCI-SCF\cite{Park2021a,Park2021} work, but could likely be improved by using more sophisticated optimization like the ones described by Yao and Umrigar or Kreplin et al.\cite{Kreplin2019,Yao2021}
}

Another strategy is to use a smaller $\epsilon_1$ value, which increases the number of determinants in the HCI wave function.
Using $\epsilon_1$, we can control the accuracy of our gradients, compared to CASSCF, and we find that we can use a relatively large value of $\epsilon_1$, i.e. a smaller number of determinants in our wave function, and still produce accurate gradients.
For the same two systems as above, we examine the convergence of the analytic gradients with respect to $\epsilon_1$, i.e. the size of the variational space.
Again we use the cc-pVDZ basis set and \cref{fig:eps1_conv} shows the results for both systems.
To contextualize the results, we show the loose, default, and tight gradient convergence tolerances used by the \texttt{Gaussian16} package\cite{Frisch2016} during geometry optimizations.
In almost all cases we find that AA rotations decrease the error relative to the CASSCF gradients, but in several instances, such as $\epsilon_1=5\cdot10^{-4}$ Ha in trans-stilbene, it does not.
Since AA rotations are not guaranteed to reduce the vHCISCF or HCISCF gradient error relative to CASSCF, it is not surprising that some cases exist where it increases the error slightly.
For all wave functions we observe a steady convergence of the gradient to the exact (CASSCF) value.
The rate of convergence varies, but shows similar behaviour for all four systems which is encouraging given the multireference nature of Sc$_2$ and Ni(edt)$_2$.
For the typical range of $\epsilon_1$ values used in research applications ($\epsilon_1=10^{-4}-10^{-5}$ Ha) the HCISCF gradients agree well with their CASSCF counterparts.
These encouraging results allow us to reduce the computational time required to calculate nuclear gradients of vHCISCF without significantly reducing their accuracy.
For the remainder of the paper, all gradients are calculated with vHCISCF and no AA rotations, unless otherwise specified.

\begin{figure*}[!htbp]
    \begin{subfigure}[t]{0.49\textwidth}
        \centering
        \includegraphics[width=\textwidth,trim={0 0 0 0.95cm},clip]{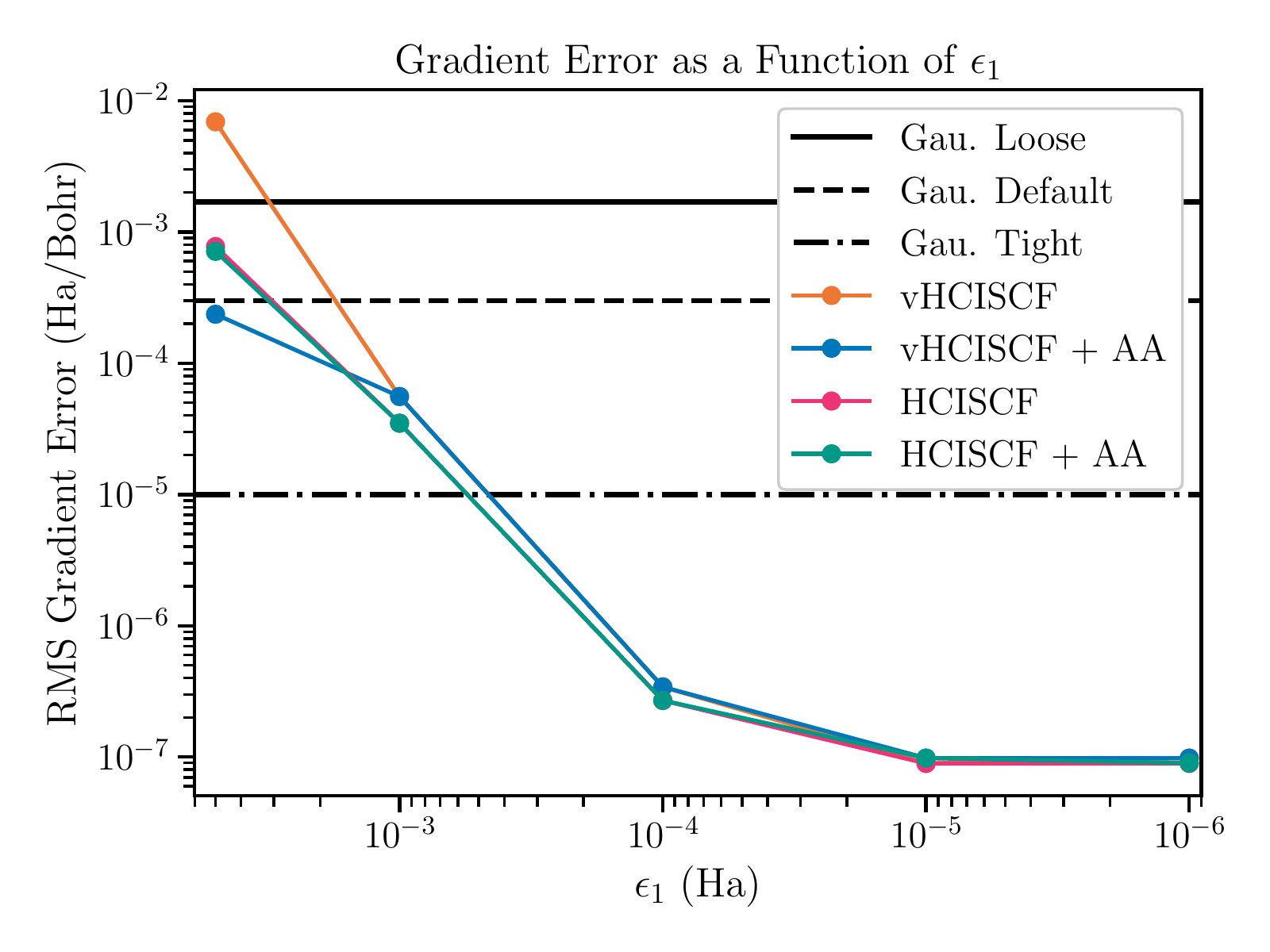}
        \caption{}
    \end{subfigure}
    \begin{subfigure}[t]{0.49\textwidth}
        \centering
        \includegraphics[width=\textwidth,trim={0 0 0 0.95cm},clip]{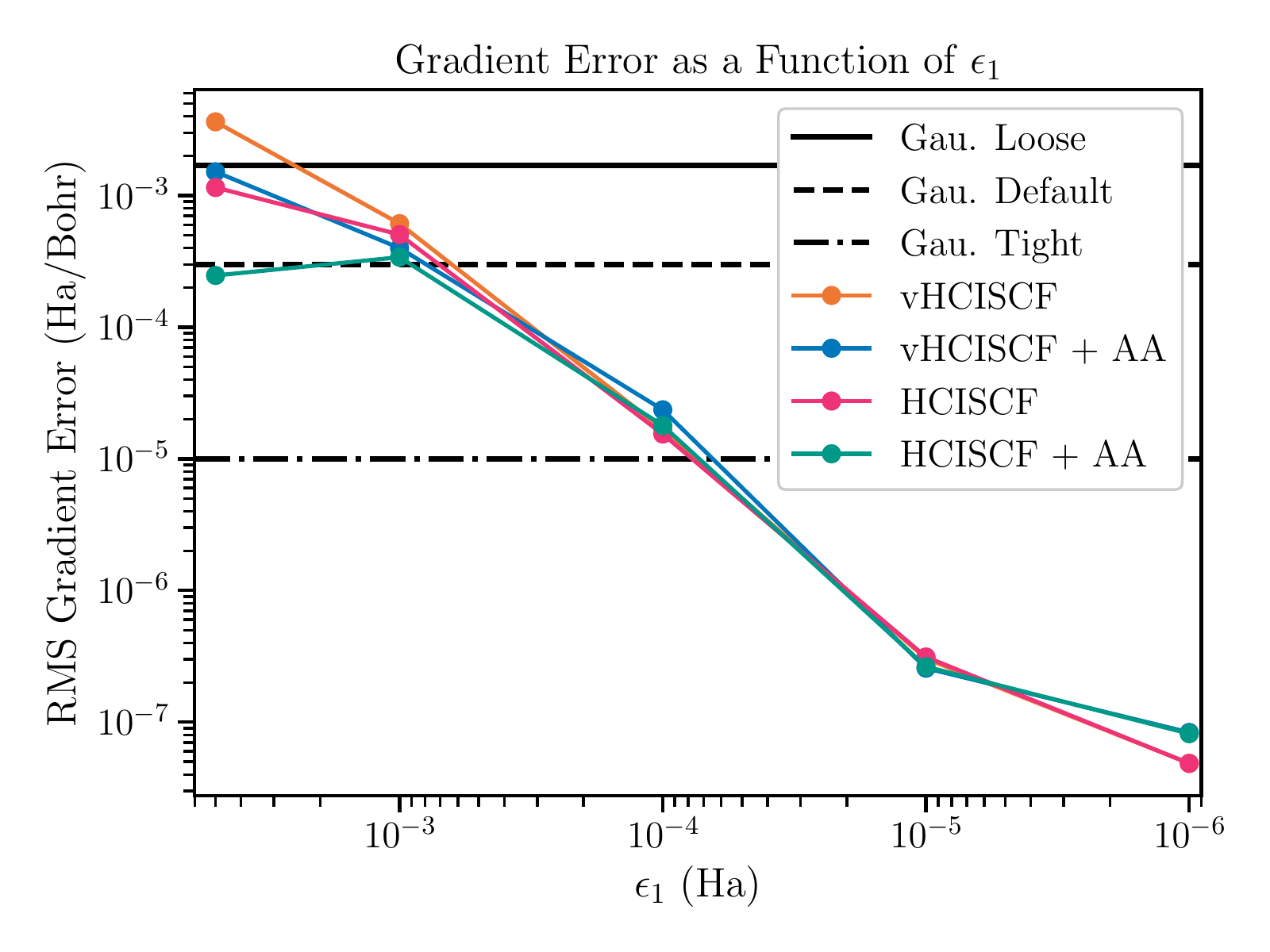}
        \caption{}
    \end{subfigure}
    \begin{subfigure}[t]{0.49\textwidth}
        \centering
        \includegraphics[width=\textwidth,trim={0 0 0 0.95cm},clip]{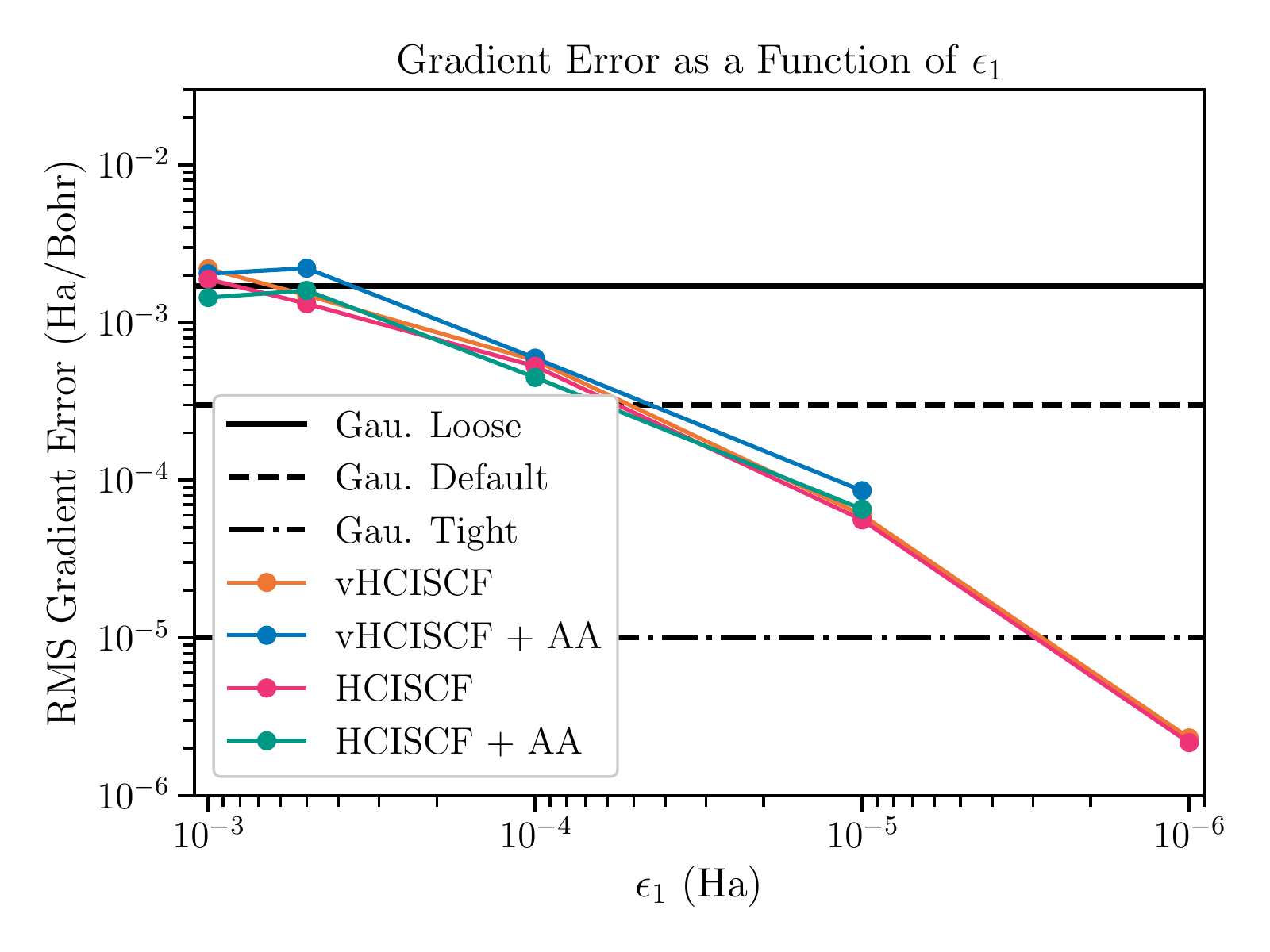}
        \caption{}
    \end{subfigure}
    \begin{subfigure}[t]{0.49\textwidth}
        \centering
        \includegraphics[width=\textwidth,trim={0 0 0 0.95cm},clip]{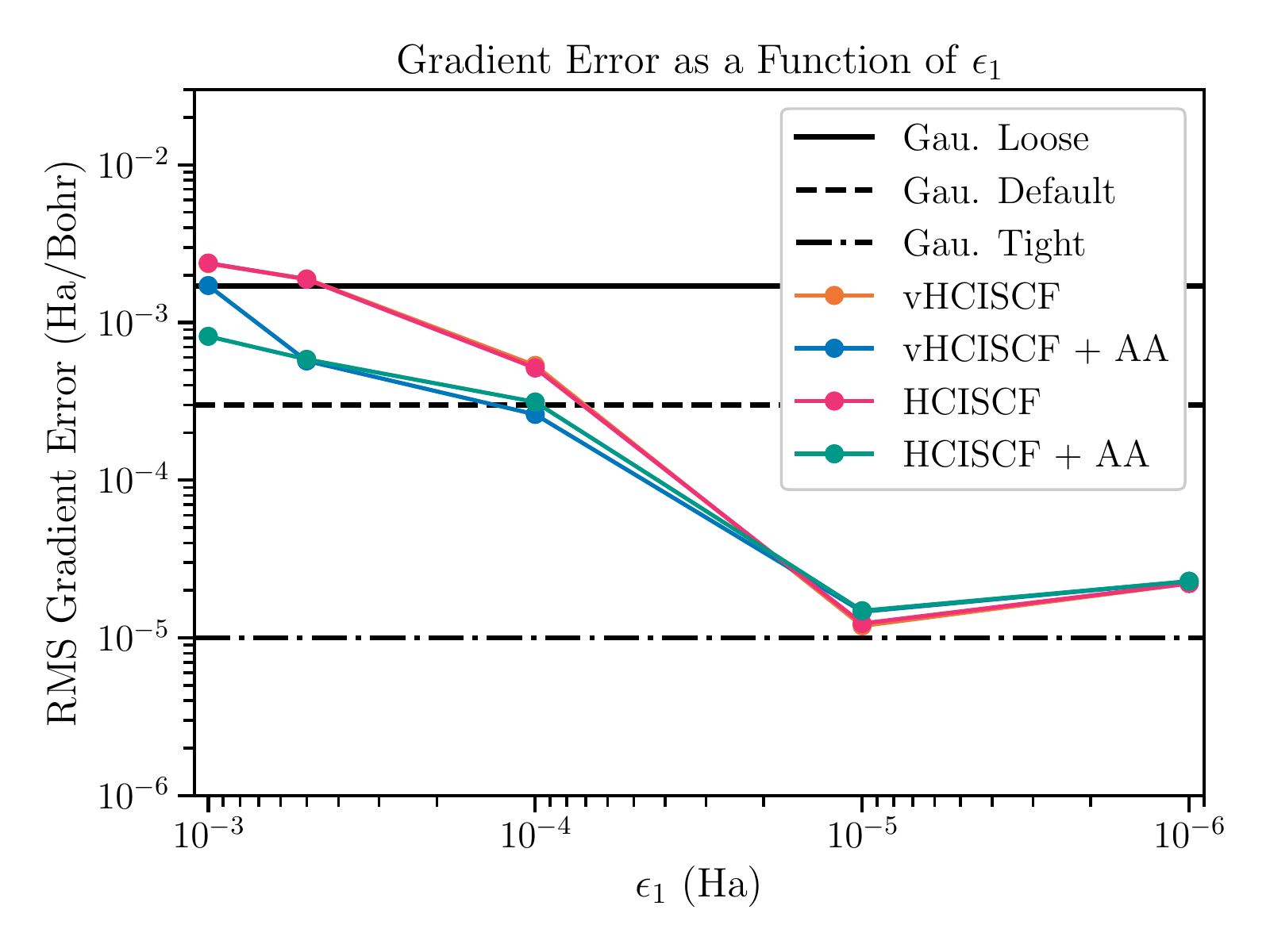}
        \caption{}
    \end{subfigure}
    \caption[Gradient error as a function of $\epsilon_1$.]{\Markup{The error in the gradient as a function of $\epsilon_1$ for several HCI-based wave functions for a) N$_2$ CAS(10e,8o)/cc-pVDZ at a bond length of 1 \AA \text{} and (b) Sc$_2$ CAS(6e,18o)/cc-pVDZ at 2.38 \AA \text{} (c) trans-stilbene CAS(14e,14o)/cc-pVDZ and (d) Ni(edt)$_2$ CAS(18e,13o)/cc-pVDZ. The values of $\epsilon_2$ are the same as in \cref{fig:grad_single_pt}. As the wave function converges to the CASSCF result the improvements from AA rotations and PT contribution make a negligible difference in the gradient. We compare the errors in the gradient to the tolerances for the root mean squared (RMS) gradient used by the \texttt{Gaussian} package during geometry optimization. In (c), vHCISCF+AA and HCISCF+AA values are omitted for $\epsilon_1=10^{-6}$ because of MCSCF convergence problems. In (d), we note that the HCISCF provides such a small improvement over vHCISCF that they are indistinguishable in this figure.}}
    \label{fig:eps1_conv}
\end{figure*}

\subsection{Geometry Optimization of Fe(PDI)}
We use two geometries reported by Ortu\~no and Cramer which they obtained using DFT: the first, "$^1$BS(1,1)", which we call \textbf{A}, optimized for the singlet spin symmetry and the second, "$^3$BS(3,1)", which we call \textbf{C}, optimized for the triplet.\cite{Ortuno2017}
Both structures were calculated using unrestricted M06-L with the def2-TZVP basis and density fitting.
In this work, we use the shorthand of (M,\textbf{X}) to describe a state with multiplicity M at geometry \textbf{X}. 

Ortu\~no and Cramer note in their paper that "although we report DFT energies, we note that they are likely quantitatively unreliable owing to significant spin contamination and unpredictable sensitivity to [multireference] character."\cite{Ortuno2017}
Given this warning, we first studied Fe(PDI) with several DFT functionals to assess how sensitive properties like energy and spin density (on the Fe atom) were to these choices.
We chose several of the Minnesota functionals (M06-2X, M06-L, and MN15)\cite{Zhao2006,Zhao2008,Yu2016} because of their success with challenging transition metal complexes. 
We also optimized geometries to test how sensitive the relaxed structures were to the choice of these functionals.

For all optimizations, we used \textbf{A} as the starting geometry for singlets and \textbf{C} as the starting geometry for triplets.
We found that for each multiplicity, the three functionals led to distinct geometries, i.e. six geometries in all.
We compare the geometries in detail in the Appendix, see \cref{tab:dft_general_comp}. 
In \cref{fig:dft_general_comparison}, we illustrate the difference in relative energy and Fe spin density at the relaxed geometries.
\Markupp{The left panel demonstrates that the spin symmetry of the ground state depends on the functional and the right panel highlights that spin density on the iron atom ranges from two to four unpaired electrons.}
We also found that the NOONs at each of these relaxed structures in \cref{tab:dft_frontier_noons} vary noticeably for different functionals.
This qualitative difference makes it challenging to determine whether singlet and triplet geometries should start from \textbf{A} or \textbf{C} in subsequent multireference calculations.
\Markup{
Throughout the study of this species, we have found that geometry optimization use DFT and vHCISCF wave functions can depend on the initial geometry which prompted us to select our initial geometries with care.
}

\begin{figure*}
    \centering
    \includegraphics[width=0.6\textwidth]{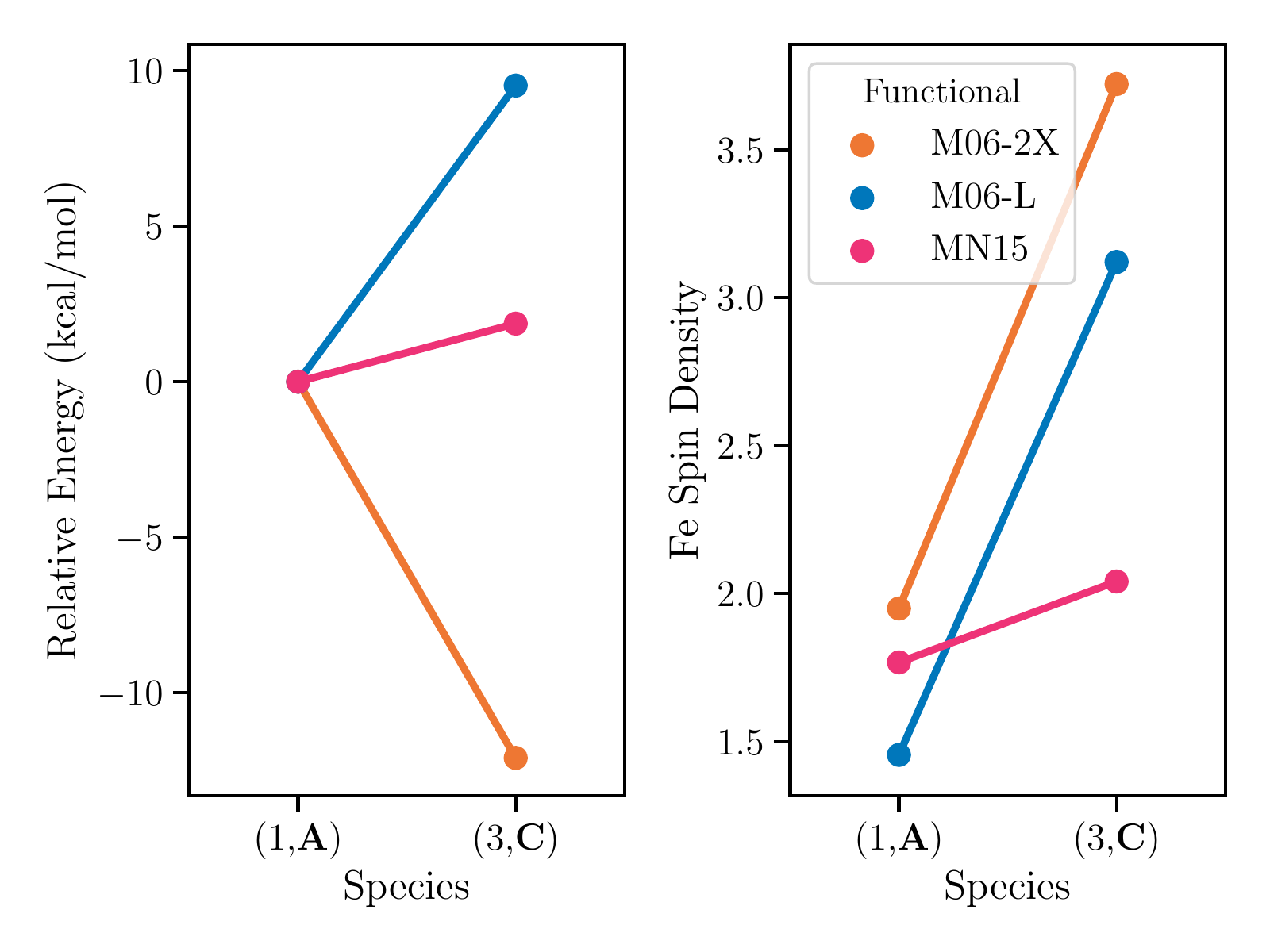}
    \caption{The effect of the functional on energy and spin density on the Fe atom for all systems studied with DFT. \Markupp{The energies and Fe spin density reported are from the relaxed geometries.} The initial coordinates for calculations are from Ortu\~no and Cramer.}
    \label{fig:dft_general_comparison}
\end{figure*}

\begin{table}[]
\begin{ruledtabular}
        \centering
    \begin{tabular}{ccccccc}
\toprule
Functional & Mult. &  Geom. &  HONO-1 &  HONO &  LUNO &  LUNO+1 \\
\midrule
\hline
    M06-2X &            1 &        \textbf{A} &   1.325 & 1.000 & 1.000 &   0.675 \\
    M06-2X &            3 &        \textbf{C} &   1.188 & 1.000 & 1.000 &   0.812 \\
     M06-L &            1 &        \textbf{A} &   1.921 & 1.000 & 1.000 &   0.079 \\
     M06-L &            3 &        \textbf{C} &   1.566 & 1.000 & 1.000 &   0.434 \\
      MN15 &            1 &        \textbf{A} &   1.617 & 1.000 & 1.000 &   0.383 \\
      MN15 &            3 &        \textbf{C} &   1.686 & 1.000 & 1.000 &   0.314 \\
\bottomrule
\end{tabular}

    \caption{The natural orbital occupation numbers (NOONs) for the systems studied with DFT as a function of functional. \Markupp{All values reported are from the relaxed geometries.}
    Here HONO is the highest occupied NO and LUNO is the lowest unoccupied NO.}
    \label{tab:dft_frontier_noons}
\end{ruledtabular}
\end{table}

\subsubsection{Selecting Initial Geometries}
\begin{figure*}
    \centering
    \includegraphics[width=0.6\textwidth]{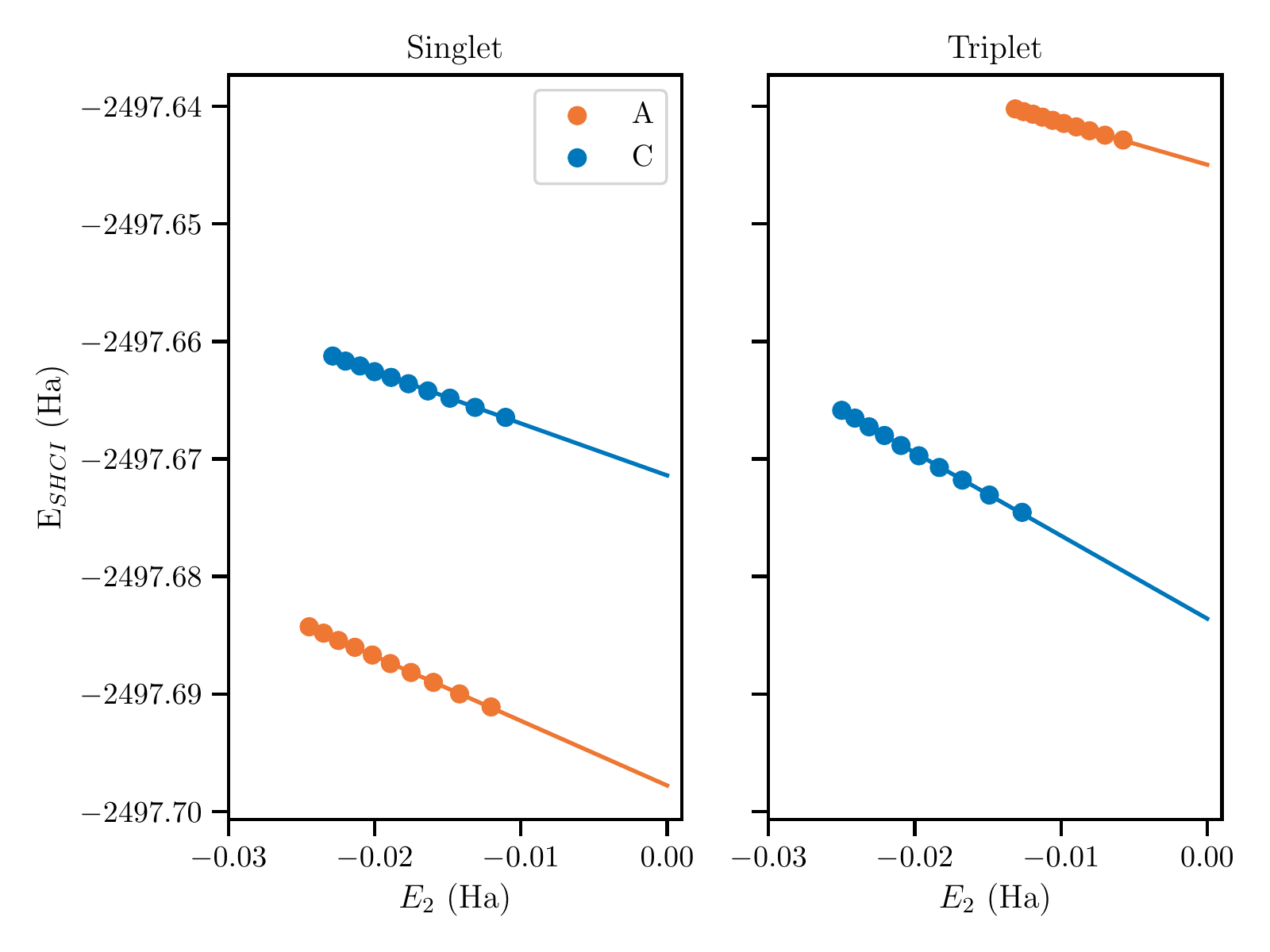}
    \caption{The extrapolated SHCI energies at the initial geometry for both the singlet and triplet spin symmetries. For the extrapolated values see \cref{tab:energy_relaxation}.}
    \label{fig:initial_extrapolation}
\end{figure*}

Since we hope to calculate accurate S-T gaps which we can qualitatively compare to experiments, it is challenging to "decide" which geometries and functionals we should use for multireference calculations.
To address this, we can use SHCI energies to determine which DFT geometry is the most stable in an unbiased manner and use these geometries in the subsequent study of the S-T gaps.
For each geometry (\textbf{A} and \textbf{C}), we used the procedure outlined above to select the active space of size (40e,40o) and then ran single point vHCISCF calculations followed by a single tight SHCI calculation.
We deliberately correlate a larger number of orbitals than would be selected by the NOON numerical thresholds mentioned earlier in order to confidently include all strongly correlated orbitals in the active space.
For the vHCISCF calculations we used a value of $\epsilon_1=7.5\cdot10^{-5}$ Ha.
Given the sensitivity of the MCSCF solutions to the input orbitals, we tested the several values of $\epsilon_1$ and found that the MCSCF orbitals converge faster than energy with respect to this parameter agreeing with previous work by several of the authors.\cite{Smith2017} 
We report this analysis in the \cref{app:mcscf_eps1_sensitivity}.
\Markup{
To ensure that our choice of $\epsilon_1$ was appropriate, we tested the use of a more accurate $\epsilon_1$ value of $5\cdot10^{-5}$ Ha and found that it produced geometries that are quite similar to those produced when using $\epsilon_1=7.5\cdot10^{-5}$.
The RMSD value between the two optimized geometries is $5.6\cdot10^{-3}$ \AA \text{}, which is less than the loose convergence criteria used by \texttt{Gaussian} ($6.7\cdot10^{-3}$ \AA).
vHCISCF wave functions using $\epsilon_1=7.5\cdot10^{-5}$ contained roughly $3.5\cdot10^6$ determinants and those using $\epsilon_1=5\cdot10^{-5}$ contained roughly $7.7\cdot10^6$.
The ability to use a larger value of epsilon reduced the time necessary to solve the CI problem from more than 1600 seconds to roughly 700 seconds.
We point out that the savings would increase when comparing to smaller values of $\epsilon_1$ and this speed-up is a rough lower bound.
All timings reported here were recorded on a single AMD Rome node with 128 cores.
}
For the final SHCI calculation we used several values of $\epsilon_1$ to extrapolate to the FCI limit, the tightest one being $10^{-5}$ Ha, and $\epsilon_2=5\cdot10^{-8}$ Ha.
We detail this extrapolation procedure in \cref{app:extrap}.
Extrapolations for singlet and triplet states at all three geometries are shown in \cref{fig:initial_extrapolation}.

\cref{fig:initial_extrapolation} indicates that \textbf{A} is the best structure to start geometry optimization from for the singlet spin symmetry and \textbf{C} for the triplet, agreeing with the DFT results using M06-L and MN15.
For completeness, we optimized the Fe(PDI) structure starting from both multiplicities at all initial geometries for a total of four vHCISCF geometry optimizations.
For all vHCISCF gradients used in geometry optimizations, we use \cref{eqn:casscf_grad} without including any AA rotations or response terms mentioned in \cref{sec:theory} making the gradients approximate.
This approach is consistent with several previous works where gradients of approximate CASSCF methods were calculated.\cite{Liu2013,Hu2015,Maradzike2017,Mullinax2019,Zimmerman2019,Park2021a,Park2021}
\Markup{In recent work, Park found little difference between ASCI-SCF geometries compared to CASSCF so we believe this approximation is justified.\cite{Park2021}
Even though the systems studied by Park are not strongly correlated, our results for Ni(edt)$_2$ suggest that even strongly correlated species follow the same trend.
}

\begin{figure*}
    \centering
    \includegraphics[width=0.6\textwidth]{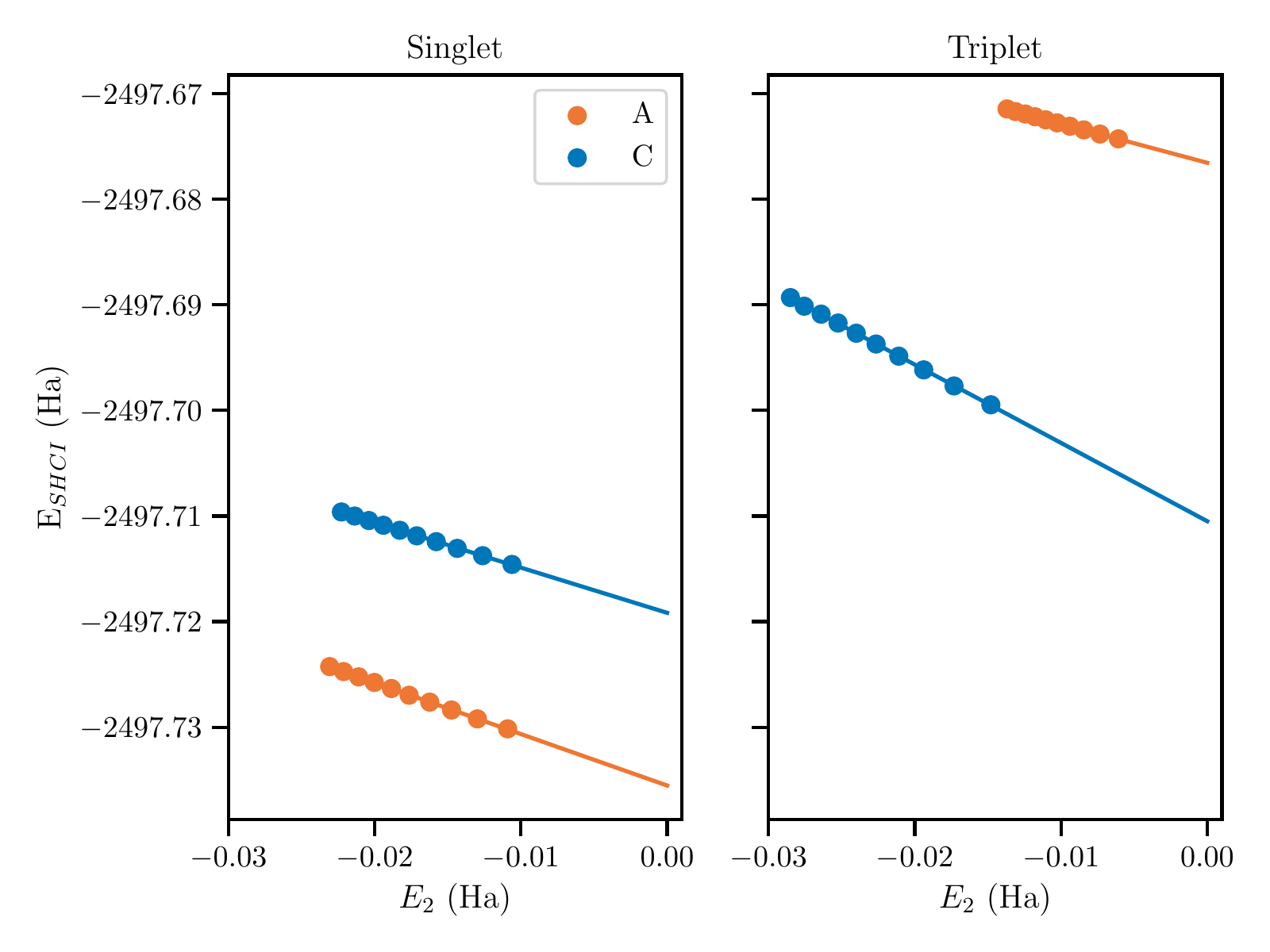}
    \caption{The extrapolated SHCI energies at the final (i.e. "relaxed") geometries for both singlet and triplet spin symmetries. For the extrapolated values see \cref{tab:energy_relaxation}.}
    \label{fig:final_extrapolation}
\end{figure*}

\subsubsection{Comparing vHCISCF Optimized Geometries}
\Markup{
In the following discussion of optimized geometries, we consider differences in two geometries to be small when they are roughly equal to the tolerance we use for convergence, $\tau_{drms}$ ($2.4 \cdot 10^{-3}$ \AA) and large when they are more than one order of magnitude larger.
}
The RMSD between initial and final geometries for all vHCISCF optimization are between 0.1-0.4 \AA \text{} indicated that the structures changed in a substantial manner and the detailed values are shown in Table \ref{tab:rmsd_effect_of_opt}.
Since this difference is orders of magnitude larger than our threshold for convergence, ($3.6\cdot 10^{-3}$ \AA), the final geometries are all distinct from their starting structures.
When comparing optimized geometries starting from different initial coordinates, the structural difference shrink for all singlet states after vHCISCF geometry optimization
while triplet states all diverge from each other. 
We show a numerical comparison of the geometries in \cref{tab:rmsd_geom_comparison}.

\begin{table}[]
\begin{ruledtabular}
    \centering
    \begin{tabular}{ccc}
\toprule
Geometries &        Singlet RMSD &        Triplet RMSD \\
\hline
\textbf{A} & $1.7 \cdot 10^{-1}$ & $1.3 \cdot 10^{-1}$ \\
\textbf{C} & $3.9 \cdot 10^{-1}$ & $3.4 \cdot 10^{-1}$ \\
\bottomrule
\end{tabular}

    \caption{Comparing the change in geometry after optimization for the singlet and triplet species. All RMSD values are in \AA \text{} and are calculated using the using the \texttt{rmsd}\cite{Kromann2020} package.}
    \label{tab:rmsd_effect_of_opt}
\end{ruledtabular}
\end{table}

\begin{table}[]
\begin{ruledtabular}
    \centering
    \begin{tabular}{cccc}
\toprule
           Geometries &             Initial &   Optimized Singlet&   Optimized Triplet\\
                    & RMSD & RMSD & RMSD\\
\hline
\textbf{A},\textbf{C} & $7.8 \cdot 10^{-1}$ & $5.2 \cdot 10^{-1}$ & $7.9 \cdot 10^{-1}$ \\
\bottomrule
\end{tabular}

    \caption{Comparing pairs of geometries at the initial, singlet optimized, and triplet optimized structures. Optimizing the singlet decreases the differences in all structures while optimizing the triplet state increases them. All RMSD values are in \AA \text{} and are calculated using the using the \texttt{rmsd}\cite{Kromann2020} package.}
    \label{tab:rmsd_geom_comparison}
\end{ruledtabular}
\end{table}

\begin{figure*}
    \centering
    \includegraphics[width=0.75\textwidth]{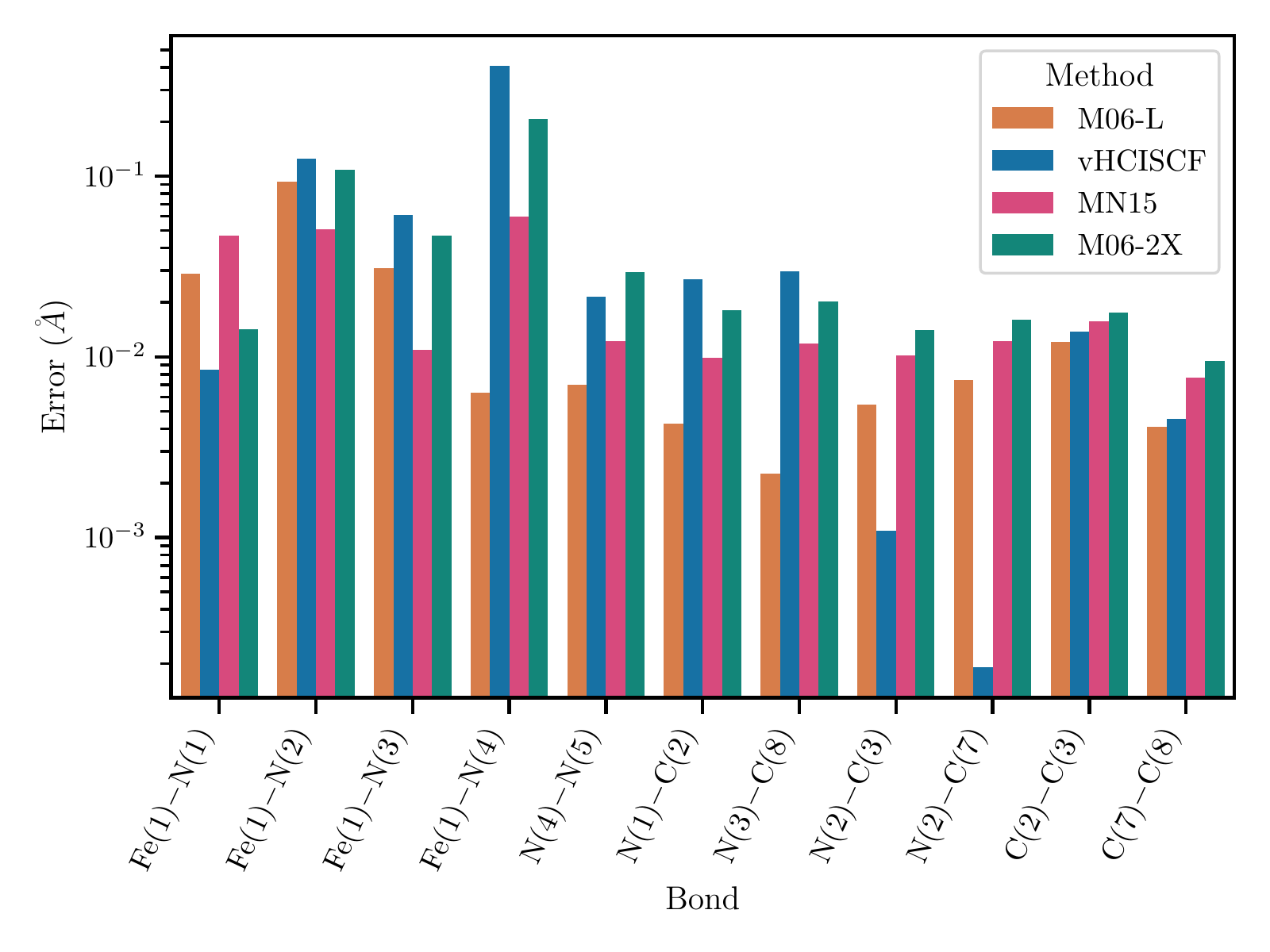}
    \caption{The errors in initial and final geometry compared to experimentally obtained bond lengths for state (1,\textbf{A}).\cite{Stieber2012} See \cref{fig:fe_pdi} for a numerically labelled diagram Fe(PDI). We report the analogous figures for other spin states and starting geometries in \cref{app:geom_comp}.}
    \label{fig:geometry_comparison_1_A}
\end{figure*}

Next we examine the differences between the experimental and theoretical structures of Fe(PDI).
For consistency, we use the same numbering scheme as Stieber et al.\cite{Stieber2012}
\cref{fig:geometry_comparison_1_A} shows error in the initial and final geometries of state (1,\textbf{A}).
To reiterate, all initial geometry were optimized with symmetry-broken DFT, while the final geometries was obtained with vHCISCF.
We compare select bond lengths from both geometries to experimental ones reported in Ref. \citenum{Stieber2012} which were obtained from crystallographic experiments.
We only report a subset of the bond lengths because (1,\textbf{A}) is only a model system for the experimental complex and they differ slightly in the composition.
Despite using multireference methods, we found that the optimized geometries did not outperform the M06-L, MN15, and M06-2X.
However, the DFT geometry optimizations use def-TZVP while the vHCISCF calculations use cc-pVDZ.
The def-TZVP basis is considerably larger than cc-pVDZ and should \textit{a priori} provide a better description of geometry.
Due to computational limitations, we were not able to run vHCISCF calculations using triple zeta basis sets.
\Markupp{We found that the time to calculate the nuclear gradients increased by nearly 50 fold while the time per MCSCF iteration increased by less than a factor of two, making the gradients the clear bottleneck to using triple zeta basis sets.
}
We note that the agreement between the DFT-optimized and experimental geometries may be fortuitous since the Fe(PDI) complex contains more alkyl groups and the experimental evidence suggests that the electronic structure is sensitive to these changes.\cite{Stieber2012}
However, a further investigation is necessary to test this claim.
The other geometry optimizations starting from other geometries and/or spin symmetry showed similar behavior and we report them in \cref{app:geom_comp}.

\textbf{}

\subsubsection{Singlet-Triplet Gaps}
Ref. \citenum{Stieber2012} argued that the ground state is a singlet with a thermally accessible triplet state based on the strong temperature dependence of NMR data for species similar to Fe(PDI).
However, the set of compounds they report in Table 2 of their paper differ by the alkyl (methyl, ethyl, or isopropyl) groups attached to C(2) and C(8) compared to Fe(PDI). 
The type of functional group noticeably impacts the extent of the temperature dependence and it is not clear if those same arguments apply to our model species.
In this work, we define thermally accessible as roughly 1 kcal/mol, i.e. "chemical accuracy".
Ref. \citenum{Ortuno2017} also suggested that the S-T gap of Fe(PDI) small and reported vertical excitation values of 2.2 kcal/mol using RASSCF and 3.4 kcal/mol using RASPT2.
Their RAS calculations correlated 22 electrons in 22 orbitals and use a mixed basis where a triple-zeta basis is used for Fe, a double-zeta basis is used for C and N, and a minimal basis is used for H.

Our calculations agree with previous experimental and theoretical work that the ground state has singlet spin symmetry, but do not indicate that there is a low-lying triplet state.
Our vertical excitation energies at \textbf{A} are roughly an order of magnitude larger than those reported by Ortu\~no and Cramer, while the initial adiabatic gap is more comparable.
During optimization, the singlet species relax between 24-30 kcal/mol, while the triplet species relax by only 17-20 kcal/mol, leading to larger S-T gaps in contrast to Ortu\~no and Cramer's expectations.
We report the S-T gaps in \cref{tab:s_t_gaps} from extrapolated SHCI calculations along with DFT and RAS calculations from Ortu\~no and Cramer\cite{Ortuno2017} for reference.

\begin{table}[]
\begin{ruledtabular}
    \centering
    \begin{tabular}{lccc}
\toprule
          Method &                          States &  Initial & Final \\
\hline
\multicolumn{4}{c}{Ortu\~no and Cramer\cite{Ortuno2017}} \\
           M06-L & (3,\textbf{C}) - (1,\textbf{A}) &      9.5 &     - \\
RASSCF (22e,22o) & (3,\textbf{A}) - (1,\textbf{A}) &      2.2 &     - \\
RASPT2 (22e,22o) & (3,\textbf{A}) - (1,\textbf{A}) &      3.4 &     - \\
\multicolumn{4}{c}{This Work} \\
  SHCI (40e,40o) & (3,\textbf{A}) - (1,\textbf{A}) &     33.2 &  37.0 \\
  SHCI (40e,40o) & (3,\textbf{C}) - (1,\textbf{A}) &      8.9 &  15.7 \\
\bottomrule
\end{tabular}
    \caption{The singlet-triplet gaps in kcal/mol at the initial (DFT-optimized) reported by Ortu\~no and Cramer\cite{Ortuno2017} and final (vHCISCF-optimized) geometries. We show the states used to calculate each gap and point out that RASSCF and RASPT2 geometries are vertical excitations at geometry \textbf{A}, while the all other initial gaps are adiabatic, i.e. at two distinct relaxed geometries.}
    \label{tab:s_t_gaps}
\end{ruledtabular}
\end{table}

\Markup{
\subsubsection{Multiple Local Minima}
During initial DFT geometries optimizations we found another triplet geometry optimization with M06-L, which we call \textbf{B}.
We found that this geometry was more similar to \textbf{A} than \textbf{C} and when compared to \textbf{A} had an RMSD value of $2.8\cdot10^{-2}$ \AA \text{} and differed in energy by roughly 1 kcal/mol.
After running both singlet and triplet vHCISCF geometries optimizations starting from \textbf{B}, we found that the singlet geometry became much closer to \textbf{A} ($8.0\cdot10^{-3}$ \AA \text{}) and had an extrapolated energy that agreed within $1.72 \cdot 10^{-5}$ Ha which is nearly the standard error ($1.56 \cdot 10^{-5}$) Ha, leading us to believe that these are the same structure.
On the other hand, after geometry optimization for triplet spin symmetry, we found that the geometry differed from the relaxed \textbf{C} geometry by $7.9\cdot10^{-1}$ \AA \text{}, but had an extrapolated energy that differs by less than 1 kcal/mol implying that there are at least two distinct stable triplet structures and that initial geometry choices are important for this species.

\begin{table}[]
    \centering
    \begin{tabularx}{\linewidth}{@{\extracolsep{\fill} } ccccc}
\hline\hline
 Mult. & Geom. & Initial & Final & Relaxation \\
 && (Ha) & (Ha) & (kcal/mol)\\
\hline
            1 &        \textbf{A} &         -2497.69779 &       -2497.73552 &                -23.67 \\
            1 &        \textbf{B} &         -2497.69678 &       -2497.73550 &                -24.29 \\
            1 &        \textbf{C} &         -2497.67140 &       -2497.71916 &                -29.97 \\
            3 &        \textbf{A} &         -2497.64496 &       -2497.67656 &                -19.83 \\
            3 &        \textbf{B} &         -2497.68006 &       -2497.70922 &                -18.30 \\
            3 &        \textbf{C} &         -2497.68359 &       -2497.71049 &                -16.88 \\
\hline\hline
\end{tabularx}

    \caption{The energy relaxation for \textbf{A}, \textbf{B}, and \textbf{C} after vHCISCF geometry optimization. The initial and final energies are the extrapolated values and all have uncertainties of roughly $1.5 \cdot 10^{-5}$ Ha.}
    \label{tab:energy_relaxation}
\end{table}
}

\section{Conclusion}
We have benchmarked nuclear gradients of HCI-type wave functions and demonstrated that like orbitals, nuclear gradients are relatively insensitive to the empirical parameter in HCI.
While the nuclear gradients are formally inexact, this error tends to lead to only minimal differences in the final optimized geometries.
If more accurate gradients or optimized structures are required, we note that the gradients are systematically improvable if more accurate gradients are required, they are systematically improvable.
As a result, accurate molecular structures can be obtained with a less computational cost and if accurate energies are desired, one can run a final more accurate SHCI calculation at this geometry.

We studied the challenging catalytic model, Fe(PDI), and used HCI-type wave functions to assess geometries obtained from DFT and to optimize new structures.
\Markupp{
Given the sensitivity of geometry optimization to the starting coordinates, HCI-based methods can unambiguously compare different geometries and determine the lowest energy structure when DFT cannot.
} 
For multireference systems, where single-reference methods, like DFT and coupled-cluster, will give ambiguous results or even fail, HCI-type wave functions are a useful alternative.
Our results agree with previous experimental and theoretical work that the ground state of the Fe(PDI) complex is a singlet.
However, we found that the S-T was larger than previously predicted and was \textit{not} thermally accessible.
As alluded to earlier, there are several factors which could explain this difference such as the missing functional groups in our model system as well as different correlated methods and basis sets.
When comparing to the experimentally obtained structure, vHCISCF performs similarly, albeit slightly worse, than the three DFT functionals studied.
Fe(PDI) is a challenging system even for accurate multireference methods and we encourage others in the community to use it as a benchmark system going forward.

\section{Data Availability Statement}
The data that support the findings of this study are openly available at \href{https://github.com/jamesETsmith/hci_nuc_gradients}{https://github.com/jamesETsmith/hci\_nuc\_gradients}.

\section{Acknowledgements}
This work utilized resources from the University of Colorado Boulder Research Computing Group, which is supported by the National Science Foundation (awards ACI-1532235 and ACI-1532236), the University of Colorado Boulder, and Colorado State University.
S.S. acknowledges funding from the NSF award CHE1800584 and the Sloan research fellowship. 
J.E.T.S. gratefully acknowledges support from a fellowship through The Molecular Sciences Software Institute under NSF Grant ACI-1547580.
J.E.T.S. was also supported by the Flatiron Institute, which is a division of the Simons Foundation.
J.L. thanks David Reichman for encouragement and support.


\appendix

\section{\label{app:extrap}Extrapolation Procedure}

All calculations used for extrapolation use $\epsilon_2=5\cdot10^{-8}$ Ha and ten different values of $\epsilon_1$ ranging from $5\cdot10^{-5}$ to $10^{-5}$ Ha.
The procedure for the extrapolation is as follows:

\begin{enumerate}
    \item Run all SHCI calculations
    \item Collect the PT correction ($E_2$) (our independent variable) and total energy ($E_{SHCI}$)
    \item Fit linear curves to $E_{SHCI}$ as a function of $E_2$
    \item Estimate the error in this extrapolated energy as the standard deviation of the curve fitting parameter representing the y-intercept.
\end{enumerate}


\section{\label{app:dft_geom_opt}DFT Geometry Optimization}
Here we provide the numerical data for \cref{fig:dft_general_comparison} along with $\expval{S^2}$ and the RMSD compared to the singlet and triplet geometries from Ortu\~no and Cramer for in \cref{tab:dft_general_comp}.

\begin{table*}[]
\begin{ruledtabular}
     \centering
     \begin{tabular}{ccccccc}
\toprule
Functional & Multiplicity & Starting Geometry &  Energy (Ha) &  Fe Spin Density &  $\ev{\hat{S}^2}$ &            RMSD (Å) \\
\hline
    M06-2X &            1 &        \textbf{A} & -2506.345333 &         1.949442 &            1.9345 & $1.1 \cdot 10^{-1}$ \\
    M06-2X &            3 &        \textbf{C} & -2506.364616 &         3.721854 &            3.9329 & $3.3 \cdot 10^{-1}$ \\
     M06-L &            1 &        \textbf{A} & -2506.768403 &         1.454747 &            1.2123 & $6.6 \cdot 10^{-4}$ \\
     M06-L &            3 &        \textbf{C} & -2506.753225 &         3.120475 &            2.9871 & $4.4 \cdot 10^{-3}$ \\
      MN15 &            1 &        \textbf{A} & -2505.529508 &         1.767367 &            1.6642 & $3.3 \cdot 10^{-2}$ \\
      MN15 &            3 &        \textbf{C} & -2505.526531 &         2.040751 &            2.5562 & $7.8 \cdot 10^{-1}$ \\
\bottomrule
\end{tabular}

     \caption{Comparing the DFT results as a function of starting geometry and functional. The RMSD values are calculated relative to broken symmetry geometries Ortu\~no and Cramer reported. All singlets are compared to their "BS(1,1)" geometry and triplets were compared to "BS(1,3)".}
     \label{tab:dft_general_comp}
\end{ruledtabular}
\end{table*}

\section{\label{app:uhf_survey}Stable UHF Solutions}
Since the UDFT and UHF calculations used to prepare orbitals for the multireference calculations were so sensitive to their optimization settings, we surveyed a large number of these settings to find the lowest energy state possible.
We emphasize that such care is needed for this species because MCSCF optimizations are sensitive to the initial orbitals for this species.
We examined six different optimization options for the DFT stage and three for the UHF stage leading to a total of 108 ways to prepare the orbitals.
In \cref{tab:uhf_most_stable}, we show a subset of these calculations where the final UHF produced a stable wave function with respect to the orbital rotation parameters.

\begin{table*}
    \centering
    \begin{tabularx}{\linewidth}{@{\extracolsep{\fill} } ccccccc}
\hline\hline
 Multiplicity & Geometry & DFT Opt. Strategy & UHF Opt. Strategy &  Fe Spin Density &  $\expectationvalue{S^2}$ &  Energy (Ha) \\
\hline
            1 &        A & Fast Newton+ADIIS &            NEWTON &  \num{-2.05e+00} &            \num{3.16e+00} & -2497.225037 \\
            1 &        A &             ADIIS &            NEWTON &  \num{-2.05e+00} &            \num{3.16e+00} & -2497.225036 \\
            1 &        A &            NEWTON &            NEWTON &   \num{2.05e+00} &            \num{3.19e+00} & -2497.224119 \\
            1 &        A & Fast Newton+ADIIS &             ADIIS &  \num{-2.01e+00} &            \num{2.69e+00} & -2497.219975 \\
            1 &        A &             ADIIS &             ADIIS &  \num{-2.01e+00} &            \num{2.70e+00} & -2497.219970 \\
            1 &        A &  Fast Newton+DIIS &            NEWTON &   \num{7.50e-02} &            \num{2.75e+00} & -2497.188196 \\
            1 &        A &              DIIS &            NEWTON &  \num{-9.79e-02} &            \num{2.75e+00} & -2497.188062 \\
            1 &        A &  Fast Newton+DIIS &             ADIIS &  \num{-3.90e-04} &            \num{2.63e+00} & -2497.187794 \\
            1 &        A &              DIIS &             ADIIS &   \num{9.10e-04} &            \num{2.61e+00} & -2497.175954 \\
            1 &        C &             ADIIS &            NEWTON &  \num{-1.98e+00} &            \num{3.60e+00} & -2497.207645 \\
            1 &        C &            NEWTON &            NEWTON &  \num{-1.78e+00} &            \num{4.16e+00} & -2497.200086 \\
            1 &        C & Fast Newton+ADIIS &            NEWTON &   \num{1.77e+00} &            \num{4.13e+00} & -2497.199389 \\
            1 &        C &  Fast Newton+DIIS &            NEWTON &  \num{-1.77e+00} &            \num{4.13e+00} & -2497.199389 \\
            1 &        C &  Fast Newton+DIIS &             ADIIS &  \num{-1.76e+00} &            \num{4.12e+00} & -2497.198453 \\
            1 &        C &            NEWTON &             ADIIS &  \num{-1.80e+00} &            \num{3.89e+00} & -2497.194584 \\
            1 &        C & Fast Newton+ADIIS &             ADIIS &   \num{1.80e+00} &            \num{3.88e+00} & -2497.194527 \\
            1 &        C &              DIIS &            NEWTON &  \num{-1.33e-02} &            \num{3.54e+00} & -2497.176719 \\
            1 &        C &              DIIS &             ADIIS &   \num{7.18e-03} &            \num{3.34e+00} & -2497.175636 \\
            3 &        A &             ADIIS &            NEWTON &   \num{2.06e+00} &            \num{3.81e+00} & -2497.206466 \\
            3 &        A &             ADIIS &             ADIIS &   \num{2.05e+00} &            \num{3.58e+00} & -2497.205565 \\
            3 &        A &              DIIS &            NEWTON &   \num{1.25e-01} &            \num{4.10e+00} & -2497.204954 \\
            3 &        A &  Fast Newton+DIIS &            NEWTON &   \num{1.88e+00} &            \num{4.61e+00} & -2497.196780 \\
            3 &        A &            NEWTON &            NEWTON &   \num{1.88e+00} &            \num{4.61e+00} & -2497.196780 \\
            3 &        C &  Fast Newton+DIIS &            NEWTON &   \num{3.76e+00} &            \num{5.42e+00} & -2497.264950 \\
            3 &        C &            NEWTON &            NEWTON &   \num{3.76e+00} &            \num{5.42e+00} & -2497.264950 \\
            3 &        C & Fast Newton+ADIIS &            NEWTON &   \num{3.76e+00} &            \num{5.42e+00} & -2497.264950 \\
            3 &        C &             ADIIS &            NEWTON &   \num{3.76e+00} &            \num{5.42e+00} & -2497.264950 \\
            3 &        C &              DIIS &            NEWTON &   \num{1.84e+00} &            \num{5.11e+00} & -2497.200823 \\
            3 &        C &              DIIS &             ADIIS &   \num{1.99e+00} &            \num{4.33e+00} & -2497.196357 \\
\hline\hline
\end{tabularx}

    \caption{A summary of the various \textit{stable} UHF solutions as a function of geometry, multiplicity, as well as SCF optimization options. The "Fast Newton" strategies are heuristic options implemented in PySCF where a second order optimization is performed in a minimal basis and then projected to the desired basis to accelerate the calculation. No stable solutions were found using DIIS in the UHF optimization.}
    \label{tab:uhf_most_stable}
\end{table*}

\section{\label{app:mcscf_eps1_sensitivity}MCSCF Sensitivity to $\epsilon_1$}

\begin{figure}
    \centering
    \includegraphics[width=0.45\textwidth]{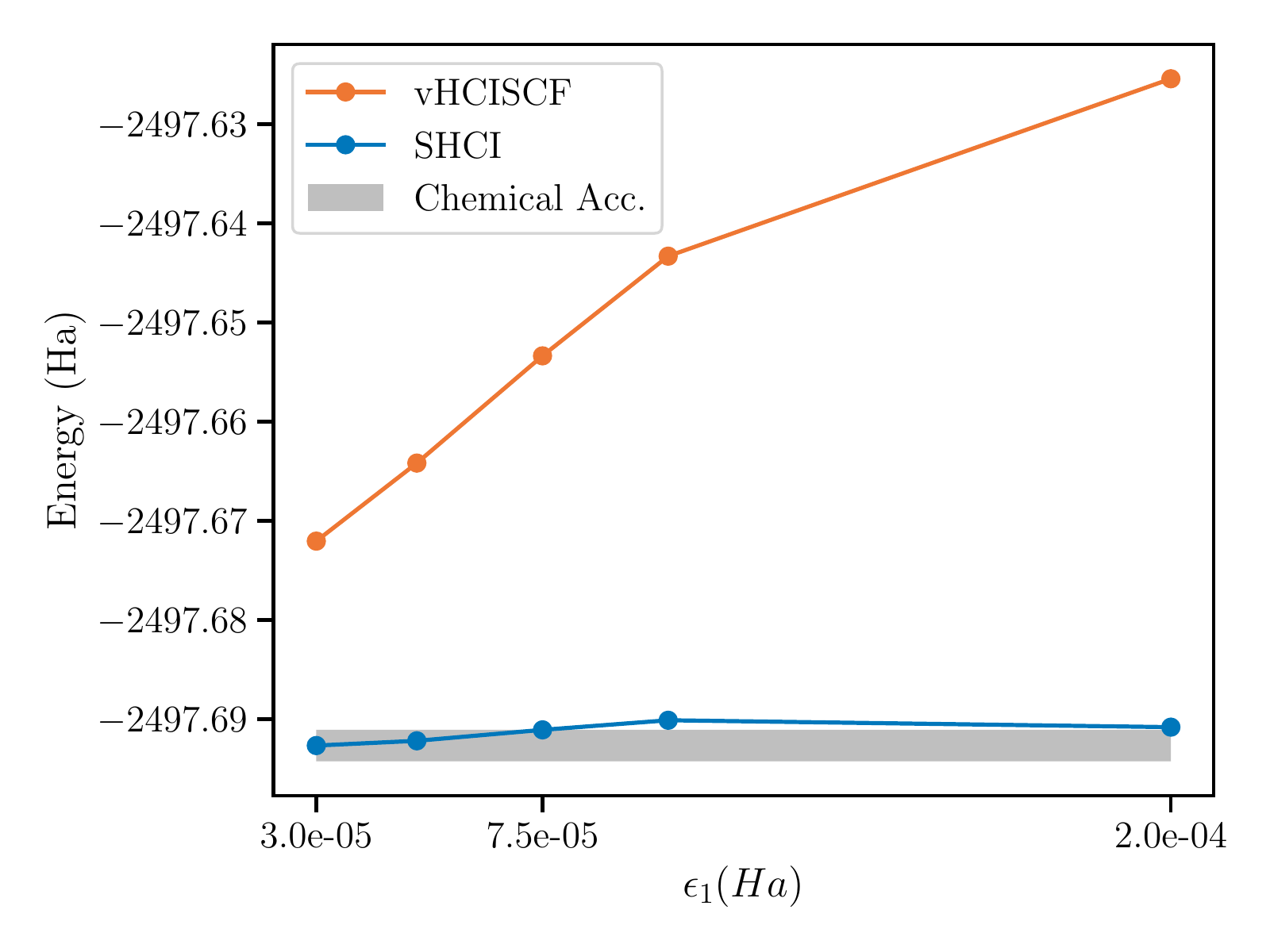}
    \caption{We report the sensitivity of the vHCISCF orbitals with respect to $\epsilon_1$. In the two step procedure we first optimize the vHCISCF orbitals and then run a final "tight" SHCI calculations using those orbitals to obtain a more accurate energy. The $\epsilon_1$ on the x-axis is for the vHCISCF calculations and indicates the accuracy/quality of the wave function with more accurate wave functions on the left. Our results show that the SHCI energy is relatively insensitive to the $\epsilon_1$ used during the vHCISCF optimization.}
    \label{fig:mcscf_eps1_sensitivity}
\end{figure}

We investigated the sensitivity of the vHCISCF orbitals to the HCI parameter, $\epsilon_1$, analogous to the analysis from Smith et al.\cite{Smith2017}
This process involves two steps: first, converging the vHCISCF orbitals with respect to several different values of $\epsilon_1$ and second, running an SHCI calculation using these orbitals with identical settings.
Despite the fact that Fe(PDI) is more multireference than the Fe(porphyrin) system studies in Ref. ~\citenum{Smith2017}, we find that the orbitals are relatively insensitive to $\epsilon_1$ and show the results in \cref{fig:mcscf_eps1_sensitivity}.

\section{\label{app:geom_comp}Geometry Comparisons}
We show the error between several theoretical results and the experimentally determined bond lengths for a subset of bond lengths near the central Fe atom.
For all states, we compare vHCISCF to M06-L and for (1,\textbf{A}), (3,\textbf{A}), and (3,\textbf{C}) we compare to MN15 and M06-2X as well, since those calculations were already performed during our earlier analysis.
In general, vHCISCF performs similarly to the ensemble of DFT results, but does not typically provide the most accurate bond lengths.


\begin{figure}
    \centering
    \includegraphics[width=0.45\textwidth]{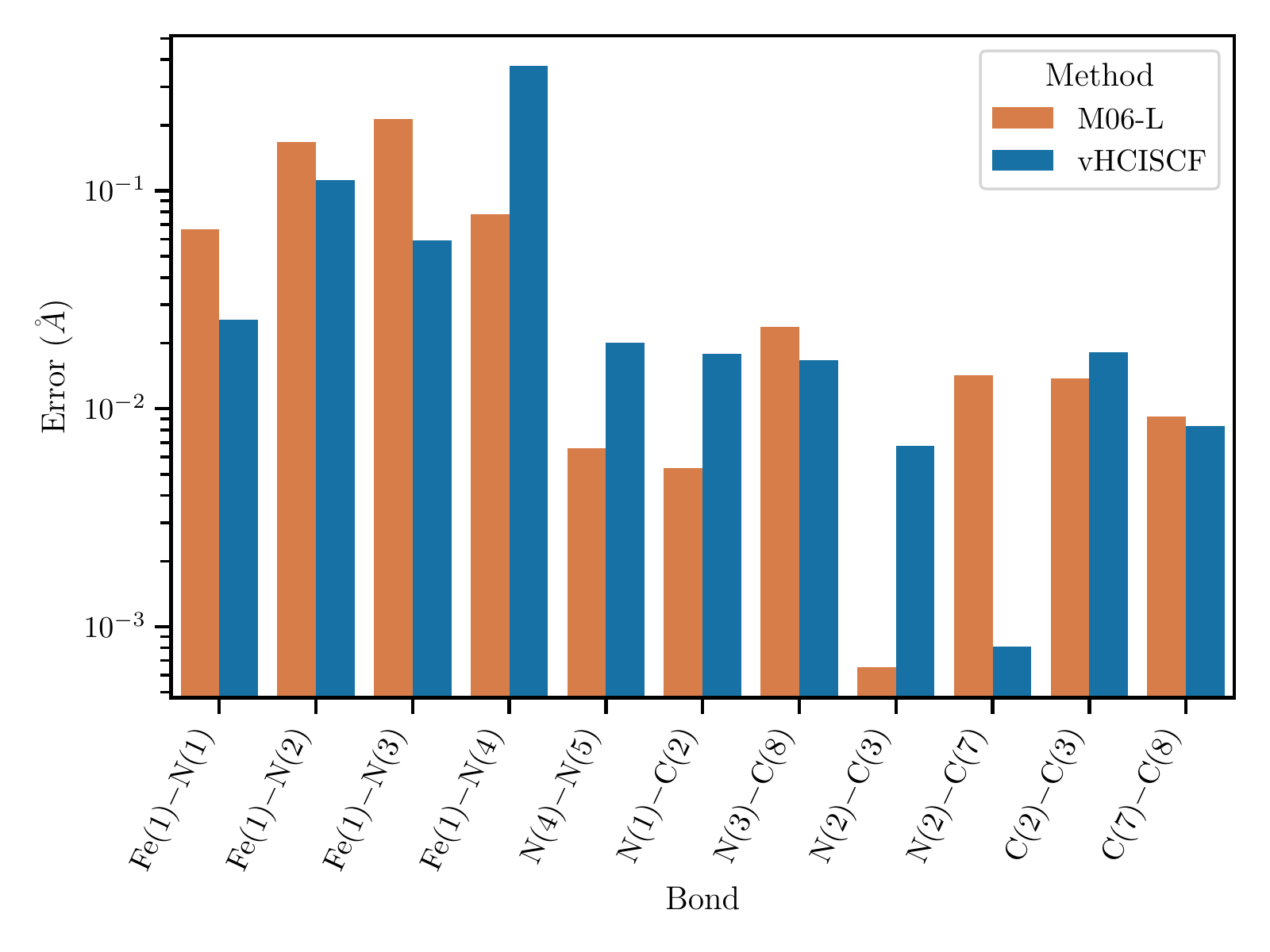}
    \caption{Comparing errors in initial and final geometry compared to experimentally obtained bond lengths for state (1,\textbf{C}).\cite{Stieber2012} }
    \label{fig:geometry_comparison_1_C}
\end{figure}

\begin{figure}
    \centering
    \includegraphics[width=0.45\textwidth]{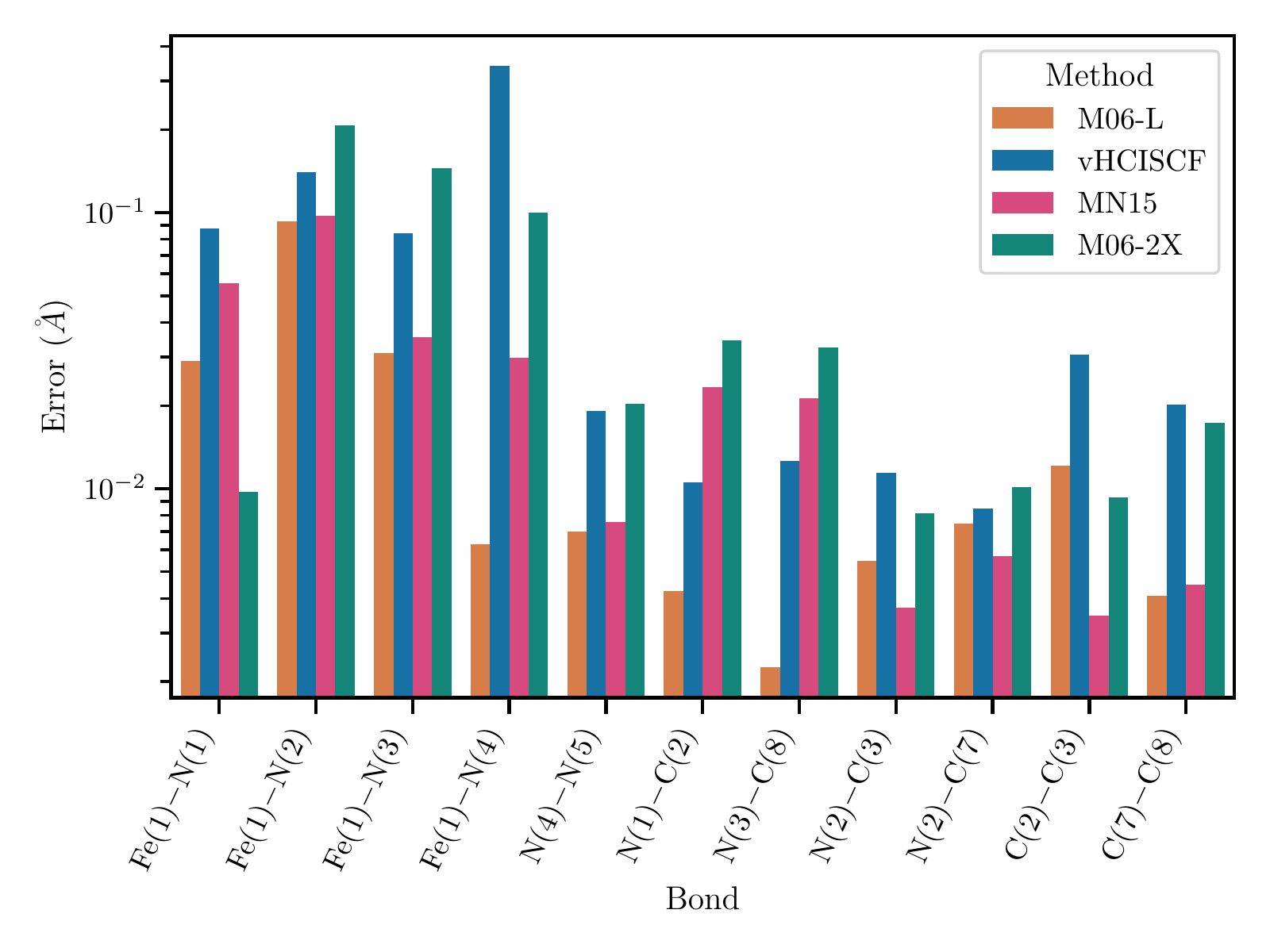}
    \caption{Comparing errors in initial and final geometry compared to experimentally obtained bond lengths for state (3,\textbf{A}).\cite{Stieber2012} }
    \label{fig:geometry_comparison_3_A}
\end{figure}


\begin{figure}
    \centering
    \includegraphics[width=0.45\textwidth]{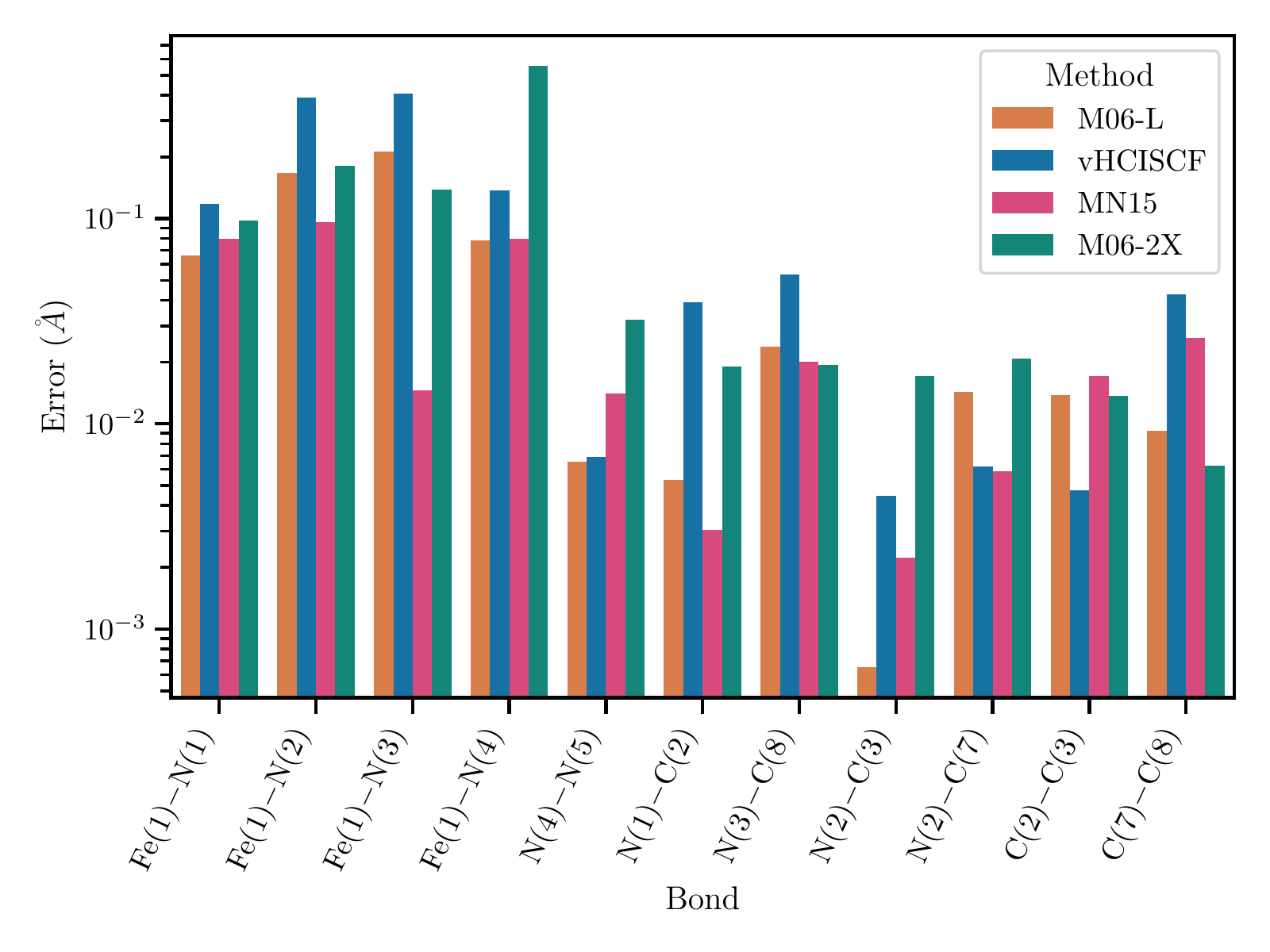}
    \caption{Comparing errors in initial and final geometry compared to experimentally obtained bond lengths for state (3,\textbf{C}).\cite{Stieber2012} }
    \label{fig:geometry_comparison_3_C}
\end{figure}


\bibliography{main}

\end{document}